\documentclass[
headsepline=false,
abstract=true,
numbers=noendperiod
]{scrartcl}
\usepackage[english]{babel}                         % fixes language-dependent typesetting standards
\usepackage[utf8]{inputenc}							% fixes source code characters: allows for umlauts and special characters through utf8
\usepackage[T1]{fontenc}							% fixes character setting: latin alphabet with T1
\usepackage{lmodern}                                % provides nicer section titles
\usepackage{fourier}								% gives a nice, slightly condensed font
\usepackage[section]{placeins}					    % prevents objects to float through sections
\usepackage{booktabs}								% produces nicer tables
\usepackage{mathtools}								% provides all kinds of math-related typesetting
\usepackage{amsmath}                                % provides various mathematical symbols
\usepackage{amssymb}								% provides various mathematical symbols
\usepackage{dsfont}                                 % provides blackboard-style double-struck letters
\usepackage{color}									% allows for coloured text
\usepackage[pdfborder={0 0 0}]{hyperref}			% set hyperlinks, without colouring
\usepackage[mathscr]{euscript}					    % gives a nicer set of script-letters ("Euler script")
\usepackage{relsize}                                % allows to enlarge mathematical symbols, e.g. \mathlarger{\sum}
\usepackage{tikz}                                   % draw tikz-pictures
\usepackage{subcaption}                             % allows captions in sub-figures
\usepackage{xfrac}                                  % provides the nice fraction \sfrac{}{}
\usetikzlibrary{decorations.pathmorphing,patterns,shapes}
\usetikzlibrary{arrows.meta}
\usepackage{multirow}                               % apparent
\usepackage{caption}                                % apparent
\usepackage{tensor}                                 % provides \tensor
\usepackage{stmaryrd}                               % provides, amongst others, the semi-simple plus sign \niplus
\usepackage[
backend=biber,
style=numeric,
citestyle=numeric-comp,
sorting=none,
doi=false,
isbn=false,
maxbibnames=99
]{biblatex}                                         % citation using biblatex
\usepackage{microtype}                              % improves layout
\DeclareMathAlphabet{\mathpzc}{OT1}{pzc}{m}{it}     % provides small script letters, e.g. \mathpzc{n}
\newcommand{\e}[1]{\operatorname{e}^{#1}}

\renewcommand{\d}{\operatorname{d}\!}
\newcommand{\tr}{\operatorname{tr}}
\newcommand{\const}{\operatorname{const}}
\newcommand{\at}[3]{\,\rule[-#1pt]{0.4pt}{#2em}\raisebox{-#1pt}{\tiny\,$#3$}}
\newcommand{\comm}[2]{\left[ #1\,, #2 \right]}
\newcommand{\acomm}[2]{\left\{ #1\,, #2 \right\}}
\renewcommand{\epsilon}{\varepsilon}
\renewcommand{\sl}{\mathfrak{sl}}
\newcommand{\isl}{\mathfrak{isl}}
\newcommand{\hs}{\mathfrak{hs}}
\newcommand{\ihs}{\mathfrak{ihs}}
\newcommand{\so}{\mathfrak{so}}
\newcommand{\R}{\mathds{R}}
\newcommand{\Msq}{\mathcal{M}^2}
\renewcommand{\S}{\mathscr{S}}
\newcommand{\C}{\mathcal{C}}
\newcommand{\Q}[4]{\,\tensor*[^{#1}_{#3}]{\mathbf{Q}}{^{#2}_{#4}}\,}
\newcommand{\StrGamma}[4]{\,\tensor*[^{#1}_{#3}]{\Gamma}{^{#2}_{#4}}\,}
\newcommand{\StrGammaB}[4]{\,\tensor*[^{#1}_{#3}]{\overline{\Gamma}}{^{#2}_{#4}}\,}
\newcommand{\ad}{\operatorname{ad}}
\newcommand{\F}[4]{\tensor*[_{2\!}]{F}{_1}\left[\left.\begin{matrix}#1\,, & #2\\ \multicolumn{2}{c}{#3} \end{matrix}\ \right| #4\right]}
\newcommand{\StrCons}[1]{#1^{\left\lfloor\!\frac{\xi}{2}\!\right\rfloor}_{\!\!j}}
\newcommand{\Vsc}{\mathcal{V}}
\newcommand{\Vb}{\bar{\mathcal{V}}}
\newcommand{\V}[4]{\tensor*[^{#1}_{#3}]{\mathbf{V}}{^{#2}_{#4}}}
\renewcommand{\c}[4]{\tensor*[^{#1}_{#3}]{c}{^{#2}_{#4}}}

\newcommand{\n}[1]{\mathpzc{n}_{\,#1}}
\numberwithin{equation}{section}
\KOMAoptions{parskip=never}
\begin{document}
\title{Scalar Fields in 3D Asymptotically Flat Higher-Spin Gravity}
\author{Martin Ammon\textsuperscript{a,}\thanks{martin.ammon@uni-jena.de} \and Michel Pannier\textsuperscript{a,}\thanks{michel.pannier@uni-jena.de} \and Max Riegler\textsuperscript{b,}\thanks{mriegler@fas.harvard.edu}}
\publishers{%
\vspace{0.6cm}
\textsuperscript{\footnotesize a\ }\begin{minipage}[t]{0.4\linewidth}
    {\footnotesize \emph{Theoretisch-Physikalisches Institut\\ Friedrich-Schiller-Universität Jena\\ Max-Wien-Platz 1, D-07743 Jena, Germany} \par}
\end{minipage}\hspace{0.5cm}
\textsuperscript{\footnotesize b\ }\begin{minipage}[t]{0.4\linewidth}
    {\footnotesize \emph{Center for the Fundamental Laws of Nature\\ Harvard University\\ 17 Oxford Street Cambridge, MA 02138, USA} \par}
\end{minipage}
}%
\date{}
\maketitle
\begin{abstract}
In this work we construct a novel associative algebra and use it to define a theory of higher-spin gravity in (2+1)-dimensional asymptotically flat spacetimes. Our construction is based on a quotient of the universal enveloping algebra (UEA) of $\isl(2,\R)$ with respect to the ideal generated by its Casimir elements, the mass squared $\Msq$ and the three-dimensional analogue of the square of the Pauli-Lubanski vector $\S$ and propose to call the resulting associative algebra $\ihs(\Msq,\S)$. We provide a definition of its generators and even though we are not yet able to provide the complete set of multiplication rules of this algebra our analysis allows us to study many interesting and relevant sub-structures of $\ihs(\Msq,\S)$. We then show how to consistently couple a scalar field to an $\ihs(\Msq,\S)$ higher-spin gauge theory.
\end{abstract}
%
% \vspace{1cm}
% \noindent{\emph{Higher-Spin Gravity, Scalar Field, Universal Enveloping Algebra, 3D Gravity, Chern-Simons}}
%
%
%
%
%
%
%\newpage
%\tableofcontents
%
%
\newpage
\section{Introduction}
The realm of (massless) higher-spin gravity in asymptotically anti-de Sitter (AdS) spacetimes experienced a great amount of interest in recent years. They have shown to be fruitful models in many different contexts: First of all, their mere existence was a surprise in the early years \cite{Vasiliev:1989qh,FRADKIN198789,Prokushkin:1998bq,Vasiliev:1999ba}. Secondly, these theories are relevant for string theory, where they appear in the tension-less limit, i.e. the limit of infinite string length \cite{Polyakov:2009pk,Polyakov:2010qs,Taronna:2010qq,Sagnotti:2010at,Sagnotti:2013bha}.  Finally, higher-spin theories may be viewed in an AdS/CFT context, in particular, they give us a chance to better understand holographic dualities. For example, in three and four spacetime dimensions, the field theories corresponding to higher-spin AdS gravity are known to be $\mathcal{W}_N$ minimal models in two dimensions \cite{Gaberdiel:2010ar,Gaberdiel:2012ku,Gaberdiel:2012uj} and $O(n)$ vector models in three dimensions \cite{Klebanov:2002ja,Sezgin:2003pt,Giombi:2012ms}, respectively .

Despite the existence of several no-go theorems regarding non-trivial, asymptotically flat higher-spin field theories \cite{PhysRev.135.B1049,PhysRev.159.1251,Aragone:1979hx,Weinberg:1980kq,Bekaert:2010hp} (see \cite{Bekaert:2010hw} for a delightful review) there has been continuous interest in the subject \cite{Bengtsson:1983pd,Bengtsson:1986kh,Metsaev:1996pd,Fotopoulos:2007nm,Manvelyan:2010jr,Fotopoulos:2010ay,Conde:2016izb,Sleight:2016xqq,Ponomarev:2016lrm,Fredenhagen:2019lsz,Metsaev:2020gmb,Skvortsov:2020pnk}. And, indeed, these no-go theorems can be circumvented in certain situations, such as in three spacetime dimensions, and lead to interesting higher-spin theories in (asymptotically) flat space \cite{Barnich:2012aw,Afshar:2013vka,Gonzalez:2013oaa,Grumiller:2014lna,Gary:2014ppa,Fuentealba:2015jma,Fuentealba:2015wza,Prohazka:2017lqb}.

Although many formulations of higher-spin gravity and their corresponding higher-spin algebras are known in the case of asymptotically AdS spacetimes \cite{Vasiliev:1989qh,FRADKIN198789,Prokushkin:1998bq,Vasiliev:1999ba,Vasiliev88,Vasiliev:1990en,Vasiliev:2003ev} (see e.g. \cite{Sorokin:2004ie,Bekaert:2005vh,Sezgin:2012ag,Didenko:2014dwa} for reviews), not so much is known about consistent formulation of higher-spin gravity and their algebras in asymptotically flat spacetime (see \cite{Bengtsson:1983pg,Bengtsson:1983pd,Bengtsson:1986kh,Metsaev:1991mt,Metsaev:1991nb} as well as \cite{Bengtsson:2014qza,Conde:2016izb,Sleight:2016xqq,Ponomarev:2016lrm} for recent developments). Previous work on higher-spin theories in 3D asymptotically flat space made ample use of taking a limit of vanishing cosmological constant of known AdS$_3$/CFT$_2$ results, e.g. on the level of the underlying Lie algebra, which translates into a certain kind of \.In\"on\"u-Wigner contraction. Taking this limit of a vanishing cosmological constant, however, can be subtle. Even at the level of a simple example like two copies of a Virasoro algebra $\mathfrak{vir}\oplus\mathfrak{vir}$ there are two possible and physically distinct contractions \cite{Bagchi:2012cy} that yield either the Galilean conformal algebra in two dimensions $\mathfrak{gca}_2$ \cite{Bagchi:2009my} or the Bondi-Van der Burg-Metzner-Sachs algebra in three dimensions $\mathfrak{bms}_3$ \cite{Bondi:1962px,Sachs:1962zza}. Even though these two algebras are isomorphic to each other they still describe two different physical situations, one where the light cone opens up completely ($\mathfrak{gca}_2$) and everything happens instantly and the other where the light cone closes up completely ($\mathfrak{bms}_3$), effectively freezing everything into place. Now, for higher-spin algebras such as e.g. $\mathcal{W}_n$ one can in principle also define and perform such contractions on the level of the Lie algebra but it is not a priori clear if there are not any other, more general contractions possible that also extend to the level of the underlying associative product structure.

With this work we intend to put the definition of a flat-space higher-spin theory on a somewhat more solid ground, in the sense that we aim for a formulation that works without explicit reference to any kind of contraction. Like in previous work on this subject we propose to define a higher-spin theory in asymptotically flat space as a Chern-Simons theory with a given underlying associative algebra structure. In AdS for example the underlying associative algebraic structure is called $\hs(\lambda)\oplus\hs(\lambda)$ \cite{Pope:1989sr,Feigin:198asr,Bordemann:1989zi,Bergshoeff:1989ns} and may be constructed as the direct sum of two copies of the UEA of $\sl(2,\R)$, quotiented by the ideals generated by their respective Casimir elements, $\lambda$ being a parametrization of these Casimir elements.

Accordingly, in the case of asymptotically flat spacetimes we propose to consider a Chern-Simons gauge theory whose gauge fields are valued in the UEA of $\isl(2,\R)$, the inhomogeneous special linear algebra, which is a semi-direct sum of $\sl(2,\R)$ and translations, $\isl(2,\R)\simeq\sl(2,\R)\niplus\R^3$\,, and construct the quotient algebra with respect to the ideals generated by the two second-order Casimir elements, the mass squared $\Msq$ and the three-dimensional analogue of the square of the Pauli-Lubanski vector $\S$.

In the AdS case, we are fortunate enough to be provided with the fundamental description of a fully interacting higher-spin theory, designed by Vasilev and collaborators \cite{Fronsdal:1978rb,Vasiliev:1989qh,FRADKIN198789,Prokushkin:1998bq,Vasiliev:1999ba}. While in its complete form being described by a genuinely complicated set of non-linear and non-local differential equations, it is well known how to linearize this theory, thereby excluding both the back-reaction of the scalar field to the higher-spin gauge fields and the interaction of the gauge fields with themselves.

Our main motivation to study such an algebraic construction is the question of how to couple a scalar field consistently to higher-spin theories in asymptotically flat spacetimes in 3D, similar to recent efforts in AdS$_3$ \cite{Mkrtchyan:2017ixk,Kessel:2018ugi,Fredenhagen:2018guf,Fredenhagen:2019hvb,Fredenhagen:2019lsz}. The interest in this coupling stems from the fact that, while three-dimensional Chern-Simons theory exhibits local degrees of freedom only on its boundary, bulk degrees of freedom may be generated by an additional scalar field. A natural starting point is inspired by the linearized version of Vasiliev's theory in three dimensions. Like in the AdS case we are dealing with a spacetime zero-form $C$, called the master field, that describes the matter content of the theory, as well as spacetime one-forms $\omega$ and $e$, called spin-connection and zuvielbein, respectively, that govern the gauge sector of the theory, i.e. describing the background geometry. Mathematically, the coupling of these fields demands some kind of star product, which in the case of Vasiliev gravity is inherently a Moyal star product that can be identified with the algebra product of $\hs(\lambda)$ \cite{Ammon:2011ua,Joung:2014qya,Korybut:2014jza,Basile:2016goq}. In our case, we claim this associative algebra to be exactly the Casimir-generated quotient of the UEA of $\isl(2,\R)$, which we call $\ihs(\Msq,\S)$.

A nice feature of our construction is that $\ihs(\Msq,\S)$ contains higher-spin algebras previously encountered in the literature, which have been derived using \.In\"on\"u-Wigner contractions, see e.g. \cite{Afshar:2013vka,Gonzalez:2013oaa,Krishnan:2013tza,Gonzalez:2014tba,Grumiller:2014lna,Riegler:2014bia,Campoleoni:2015qrh,Campoleoni:2016vsh,Ammon:2017vwt}, as well as all possible finite-spin truncations $\isl(N,\R)$. 

The structure of this paper is as follows. Section \ref{sec_review} provides a brief review of certain aspects of three-dimensional higher-spin AdS gravity that are conceptually relevant for this work. We explain how to study the propagation of scalar fields in a higher-spin gauge background in a linearized manner, and thereby provide the guiding principles for our study in asymptotically flat space.

In section \ref{sec_ihsConstruction} we lay the mathematical foundations to describe the coupling of scalar fields to higher-spin extended, asymptotically flat spacetimes. We construct the UEA of $\isl(2,\R)$ in the language of formal products, introduce the quotient algebra through formal identification of Casimir elements and define the algebra $\ihs(\Msq,\S)$ in terms of descendant generators, based on a classification of highest-weight generators. We furthermore study particular sub-structures of $\ihs(\Msq,\S)$. One is related to the algebra structure already known in the literature as a contraction from $\hs(\lambda)\oplus\hs(\lambda)$, another one seems to be the physically significant sub-structure when coupling a scalar field to an asymptotically flat higher-spin background.

In section \ref{sec_Inonu_Wigner} we give a short overview of how the algebra $\ihs(\Msq,\S)$ is related to its AdS companion $\hs(\lambda)$ via an \.In\"on\"u-Wigner contraction. The contraction prescription naturally emerges from the classical contraction of $\sl(2,\R)\oplus\sl(2,\R)$ to $\isl(2,\R)$ when considered in a formal-product representation. However, it turns out that a larger algebra than $\hs(\lambda)\oplus\hs(\lambda)$, namely a quotient of $\mathcal{U}(\sl(2,\R)\oplus\sl(2,\R))$ is necessary for a consistent contraction.

Section \ref{sec_Vasi} describes a flat-space analogue of the linear coupling of a matter field to the gauge sector in AdS. We identify an appropriate Lie-subalgebra of $\ihs(\Msq,\S)$ and derive the equations of motion for a propagating matter field, inspired by linearized Vasiliev gravity in three-dimensional AdS.

Finally, in section \ref{sec_dynamics} we couple a scalar field to a background with (constant) higher-spin charges. We then generalize this approach by including charges with arbitrary spin and present the respective generalization of the Klein-Gordon Equation for a scalar field propagating in this higher-spin background.
\section{Review: Scalar Fields in Higher-Spin AdS}\label{sec_review}
The following section is intended to give a brief overview over some of the basics of three-dimensional asymptotically AdS higher-spin gravity coupled to a scalar field. We will closely follow \cite{Ammon:2011ua}, since one goal of this work is to adapt the approach taken in this work to the flat-space case.
\subsection{Chern-Simons Formulation of 3d Gravity and Higher Spins}
It is a well-known fact that three-dimensional gravity may be recast into a Chern-Simons gauge theory \cite{Witten:1988hc} with a given gauge group, depending on the isometries of the spacetime in question. In particular, for AdS with isometry algebra $\sl(2,\R)\oplus\sl(2,\R)\simeq\so(2,2)$ the Chern-Simons gauge fields $A$ and $\bar{A}$, which are spacetime one-forms, take values in two copies of $\sl(2,\R)$, respectively. The action is given by
\begin{align}
    S[A,\bar{A}]=S_{\text{\tiny CS}}[A]-S_{\text{\tiny CS}}[\bar{A}]\,, && \text{with} && S_{\text{\tiny CS}}[A]=\frac{k}{4\pi}\int\!\left\langle A\wedge \d A+\frac{2}{3}A\wedge A\wedge A\right\rangle\,,
\end{align}
and is equivalent to the Einstein-Hilbert action (modulo boundary terms) in the metric formulation; the so-called Chern-Simons level $k$ is related to Newton's constant $G$ by $k=l/(4G)$, where $l$ denotes the AdS radius and $\langle\,,\rangle$ denotes an invariant bilinear form of the gauge algebra. Furthermore, the gauge fields are subject to the flatness conditions
\begin{subequations}
\begin{align}
    \d A+A\wedge A&=0\,,\\
    \d \bar{A}+\bar{A}\wedge \bar{A}&=0\,,
\end{align}
\end{subequations}
which are equivalent to the Einstein equations in the metric formulation. The passage from gauge-field formulation to metric formulation is easily done by decomposing the gauge field into vielbein and spin connection and identifying the metric with the trace over the vielbein squared. 

One interesting feature of a CS formulation is the simple passage to higher-spin gravity: replacing the symmetry algebra by $\sl(N,\R)\oplus\sl(N,\R)$ introduces fields of spin $s=3,\dots,N$ coupled to the classical gravity field and it is even possible to take the limit $N\rightarrow\infty$, where one encounters $\hs(\lambda)\oplus\hs(\lambda)$ as the higher-spin symmetry algebra \cite{Bergshoeff:1989ns,POPE1990191,Ammon:2011nk,Ammon:2012wc}. The latter can be constructed from the universal enveloping algebras of the two copies of $\sl(2,\R)$,
\begin{align}
    \hs(\lambda)=\frac{\mathcal{U}(\sl(2,\R))}{\left\langle\C_{\text{\tiny AdS}}\right\rangle}\,, && \text{where} && \C_{\text{\tiny AdS}}=\frac{\lambda^2-1}{4}\,,
\end{align}
and its structure constants\footnote{Note that this nomenclature does not distinguish the \emph{associative} algebra $\hs(\lambda)$ (equipped with a product) from the \emph{Lie} algebra $\hs(\lambda)$ (equipped with a bracket); in the latter case we would write the definition
\begin{align}
    \mathds{C}\oplus\hs(\lambda)=\frac{\mathcal{U}(\sl(2,\R))}{\left\langle\C_{\text{\tiny AdS}}\right\rangle}\,.
\end{align}
However, this shouldn't bother us too much because, obviously, we simply have to identify the Lie bracket with the commutator w.r.t. the associative product.} are known.
Here, $\left\langle\C_{\text{\tiny AdS}}\right\rangle$ compactly denotes the ideal generated by a parametrization of the second-order Casimir element of $\sl(2,\R)$, usually written $\left\langle\C_{\text{\tiny AdS}}-\mu\mathds{1}\right\rangle$.

Like in the spin-2 case i.e. usual 3d Einstein gravity one can also define a precise map from the higher-spin Chern-Simons gauge fields to a metric formulation by appropriately identifying the (zu)vielbein. Let us illustrate this is for the case of the $\sl(3,\R)\oplus\sl(3,\R)$ gauge algebra. The field content in a metric-like formulation then consists of the metric $g_{\mu\nu}$ and a spin-3 field $\phi_{\mu\nu\rho}$, which can be obtained using via,
\begin{subequations}
\begin{align}
  g_{\mu\nu}&\sim \tr( e_\mu e_\nu)\sim\tr\left( (A_\mu - \bar{A}_\mu) (A_\nu - \bar{A}_\nu)\right)\,,\\
  \phi_{\mu\nu\rho}&\sim \tr( e_\mu e_\nu e_\rho )\,,
\end{align}
\end{subequations}
where the trace is taken in the fundamental representation of $\sl(3,\R)$. The consistent coupling of the spin-3 field to the metric can be worked out, see e.g. \cite{Campoleoni:2012hp,Fujisawa:2012dk,Fujisawa:2013lua,Fujisawa:2013ima,Fredenhagen:2014oua}; for an exact map to all orders in the  spin-3 field see \cite{Fredenhagen:2014oua}.
\subsection{Linearized Vasiliev Gravity}\label{subsec_rev_linearisation}
An interacting theory equipped with a higher-spin gauge symmetry is given by a model developed by Vasiliev and collaborators \cite{Fronsdal:1978rb,Vasiliev:1989qh,FRADKIN198789,Prokushkin:1998bq,Vasiliev:1999ba}. Its field content consists of a spacetime one-form $W$ (containing the gauge sector of the theory) as well as two spacetime zero-forms $B$ and $S$ (taking care of vacuum/matter contributions and internal symmetries, respectively), subject to a set of non-linear equations. Key ingredient is a Moyal star-product ``$\star$'' acting on functions of (deformed) oscillator variables. All fields depend on these oscillator variables as well as on spacetime coordinates and several auxiliary elements.

We will not dive too deep into the formulation of Vasilev's theory but rather give a short sketch of its linearization. The $B$-field can be expanded around its vacuum value $\nu$ to linear order,
\begin{align}
    B=\nu+\C\,,
\end{align}
while the field $S$ just ensures correct dependencies of the fields on the various variables and does not play a vital role here. The relevant equations of motion are
\begin{subequations}
\begin{align}
    \d W&=W\wedge_{\!\star} W\,,\\
    \d \C &=W\star\C-\C\star W\,,
\end{align}
\end{subequations}
the first equation governing the gauge sector of the theory, while the second one describes the coupling of matter to the background geometry. Due to their linear nature, these equations are free of self-interactions of matter fields amongst themselves as well as back-reaction of matter fields on gravitational fields.

Next, we decompose the matter field into a dynamical and an auxiliary part with the help of a parameter $\psi_2$ that belongs to a pair of anti-commuting Clifford elements $\psi_{1/2}$ that square to one:
\begin{align}
    \C=\C^{\text{\tiny (aux)}}+\C^{\text{\tiny (dyn)}}\psi_2\,,
\end{align}
where the new fields do not depend on $\psi_2$ anymore. It turns out that $\C^{\text{\tiny (aux)}}$ can be consistently set to zero since it does not carry any degrees of freedom at all.

We can impose a further decomposition of those fields using the projection operators
\begin{align}
    \mathcal{P}_{\pm}=\frac{1\pm\psi_1}{2}\,.
\end{align}
The decomposition introduces gauge fields $A$ and $\bar{A}$ governing the gravitational sector of the theory as well as scalar matter fields $C$ and $\bar{C}$:
\begin{subequations}
\begin{align}
    W&=-\left(\mathcal{P}_+ A+\mathcal{P}_- \bar{A}\right)\,,\\
    \C^{\text{\tiny (dyn)}}&=\mathcal{P}_+ C +\mathcal{P}_- \bar{C}\,.
\end{align}
\end{subequations}
These new fields obey
\begin{subequations}
\begin{align}
    \d A+A\wedge_{\!\star}A&=0\,,\label{AdS_flatness}\\
    \d C+A\star C-C\star\bar{A}&=0\label{AdS_eom}\,,
\end{align}
\end{subequations}
together with a barred equation, respectively. Clearly, the first equation is exactly the flatness condition we already encountered in the Chern-Simons description of gravity (written in terms of the Moyal star product), while the second equation describes the coupling of the matter content of the theory to its gauge sector.

We already mentioned earlier that the algebra structure $\hs(\lambda)$ describing an infinite tower of massless fields of arbitrary (integer) spin is well-known and the gauge fields $A$ and $\bar{A}$ take values in two copies of that algebra. It is now tempting to replace the Moyal product by the associative product of $\hs(\lambda)$, known as \emph{lone-star product}; indeed, this has been proposed in \cite{Ammon:2011ua} and was proven in \cite{Joung:2014qya,Korybut:2014jza,Basile:2016goq}. One should notice the significance of these considerations: it is the associative product that describes interaction.

At this point it is instructive to discuss a particular property of the theory. Namely, it splits into a holomorphic and an anti-holomorphic sector, which, from the viewpoint of Vasiliev gravity is due to the presence of the parameter $\psi_1$ from which we build the projection operators, whereas from the viewpoint of higher-spin algebras it is due to the fact that we consider a direct sum of two copies of $\hs(\lambda)$, possibly with different values of $\lambda$. This is, however, not the only possibility to construct a higher-spin algebra within the latter point of view: alternatively, one may take the universal enveloping algebra of the direct sum $\sl(2,\R)\oplus\sl(2,\R)$, which allows for products of holomorphic and anti-holomorphic elements, as well, and yields an infinite wedge of higher-spin generators, rather than two infinite towers. This larger algebra structure will be of importance later.
\subsection{Generalized Klein-Gordon Equation}
With the considerations of the previous sub-sections we can consider the propagation of a matter field in a particular geometry in the following. First, let the generators of $\hs(\lambda)$ be $\mathcal{V}^s_m$, where $s\ge 1$ and $|m|\le s-1$. The associative product is given by \cite{POPE1990191,Gaberdiel:2011wb}
\begin{align}
    \mathcal{V}^s_m\star \mathcal{V}^t_n&=\sum_{u=0}^{s+t-|s-t|-2} g^{st}_u (m,n) \mathcal{V}^{s+t-1-u}_{m+n}\,,
\end{align}
with the structure constants defined in (\ref{lone-star_SC}) and (\ref{lone-star_N}). One may expand all fields within this algebra,
\begin{subequations}
\begin{align}
    C&=\sum_{s=1}^\infty\sum_{|m|\le s-1}C^s_m \mathcal{V}^s_m\,,\\
    A&=\sum_{s=2}^\infty\sum_{|m|\le s-1}A^s_m \mathcal{V}^s_m\,.
\end{align}
\end{subequations}

As a first case, consider the classical (spin two) AdS vacuum in Fefferman-Graham gauge given by the gauge fields
\begin{subequations}
\begin{align}
    A&=\e{\rho}\mathcal{V}^2_1 d\!z+\mathcal{V}^2_0 d\!\rho\,,\\
    \bar{A}&=\e{\rho} \mathcal{V}^2_{-1}d\!\bar{z}-\mathcal{V}^2_0 d\!\rho\,,
\end{align}
\end{subequations}
that fulfil the flatness condition (\ref{AdS_flatness}) and in the metric formulation correspond to
\begin{align}\label{metric}
    d\!s^2=d\!\rho^2+\e{2\rho} d\!zd\!\bar{z}\,.
\end{align}
Plugging those gauge fields together with the expansion of $C$ into the equation of motion (\ref{AdS_eom}) yields a closed system of coupled first-order partial differential equations for the coefficients $C^s_m$. It turns out that this system can be reduced to a Klein-Gordon equation for the coefficient in front of the unit element in $C$,
\begin{align}\label{rev_class_KG}
    \left(\Box^{(2)}-m^2\right)C^1_0=0\,,
\end{align}
where the mass is given by $m\equiv\lambda^2-1$ and
\begin{align}
    \Box^{(2)}=\frac{1}{\sqrt{-g}}\partial_{\!\mu}\left(\sqrt{-g}g^{\mu\nu}\partial_{\!\nu}\right)=\partial^2_{\!\rho}+2\partial_{\!\rho}+4\e{-2\rho}\partial_z\partial_{\bar{z}}
\end{align}
denotes the Klein-Gordon operator on the background (\ref{metric}). All remaining components $C^s_m$ can then be expressed in terms of $C^1_0$.

A chiral spin-$s$ deformation of the gauge fields may be introduced by adding a spin-$s$ highest-weight generator together with a constant coupling $\eta$:
\begin{subequations}\label{chiral_deform}
\begin{align}
    A&=\e{\rho}\mathcal{V}^2_1 d\!z-\eta\e{(s-1)\rho}\mathcal{V}^s_{s-1}d\!\bar{z}+\mathcal{V}^2_0 d\!\rho\,,\\
    \bar{A}&=\e{\rho} \mathcal{V}^2_{-1}d\!\bar{z}-\mathcal{V}^2_0 d\!\rho\,.
\end{align}
\end{subequations}
The same steps as above then lead to a generalized Klein-Gordon equation
\begin{align}
    \left(\Box^{(2)}+(-1)^s 4\eta\e{-2\rho}\partial^s_{\!z}-m^2\right)C^1_0=0\,.
\end{align}
\section{Construction of the Associative Algebra \texorpdfstring{$\ihs(\Msq,\S)$}{ihs(M\texttwosuperior,S)}}\label{sec_ihsConstruction}
As we have seen in the previous section, one of the crucial ingredients to the description of the coupling of matter fields to (higher-spin) gravitational fields is an associative product. In this section we construct such an associative product using a quotient of the UEA of $\isl(2,\R)$, which eventually also provides us with a novel higher-spin Lie algebra structure.
\subsection{General Algebra Structure in a Formal-Product Basis}\label{subsec_genStr}
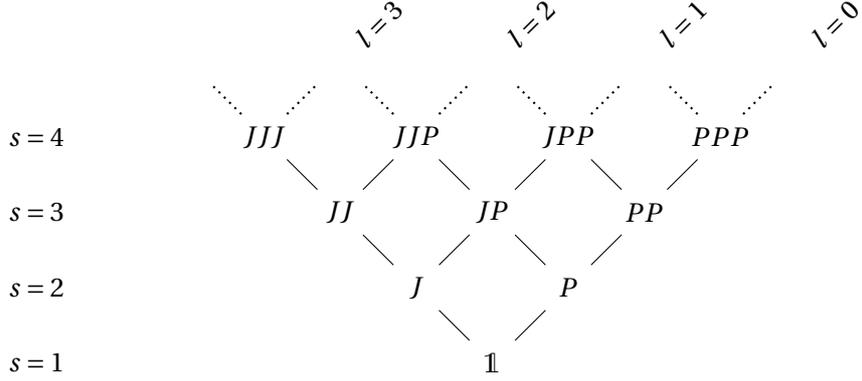
\begin{figure}[h]
\begin{center}
\begin{tikzpicture}
%\draw[fill=lightgray,color=lightgray,rounded corners,rotate=45](-0.4,-0.4) rectangle (7,0.5);
%
\node at(0,0){$\mathds{1}$};
\node at(-1,1){$J$};
\node at(1,1){$P$};
\node at(-2,2){$JJ$};
\node at(0,2){$JP$};
\node at(2,2){$PP$};
\node at(-3,3){$JJJ$};
\node at(-1,3){$JJP$};
\node at(1,3){$JPP$};
\node at(3,3){$PPP$};
\draw[-](-0.3,0.3)--(-0.7,0.7);
\draw[-](0.3,0.3)--(0.7,0.7);
\draw[-](-1.3,1.3)--(-1.7,1.7);
\draw[-](1.3,1.3)--(1.7,1.7);
\draw[-](-2.3,2.3)--(-2.7,2.7);
\draw[-](2.3,2.3)--(2.7,2.7);
\draw[dotted,thick](-3.3,3.3)--(-3.7,3.7);
\draw[dotted,thick](3.3,3.3)--(3.7,3.7);
\draw[-](-0.7,1.3)--(-0.3,1.7);
\draw[-](0.3,2.3)--(0.7,2.7);
\draw[dotted,thick](1.3,3.3)--(1.7,3.7);
\draw[dotted,thick](-0.7,3.3)--(-0.3,3.7);
\draw[-](-1.7,2.3)--(-1.3,2.7);
\draw[dotted,thick](-2.7,3.3)--(-2.3,3.7);
\draw[-](0.7,1.3)--(0.3,1.7);
\draw[-](-0.3,2.3)--(-0.7,2.7);
\draw[dotted,thick](-1.3,3.3)--(-1.7,3.7);
\draw[dotted,thick](0.7,3.3)--(0.3,3.7);
\draw[-](1.7,2.3)--(1.3,2.7);
\draw[dotted,thick](2.7,3.3)--(2.3,3.7);
\node at(-6,0){$s=1$};
\node at(-6,1){$s=2$};
\node at(-6,2){$s=3$};
\node at(-6,3){$s=4$};
\node[rotate=45] at(4.5,4.5){$l=0$};
\node[rotate=45] at(2.5,4.5){$l=1$};
\node[rotate=45] at(0.5,4.5){$l=2$};
\node[rotate=45] at(-1.5,4.5){$l=3$};
\end{tikzpicture}
\caption{The structure of $\mathcal{U}(\isl(2,\R))$ can be schematically depicted as an infinite wedge. Here we omitted the respective mode indices just focusing on the different formal powers of $J$- and $P$-generators.}
\label{uea_wedge}
\end{center}
\end{figure}

We start by building the universal enveloping algebra of $\isl(2,\R)$, which is given by generators of rotations $J_n$ and translations $P_n$, where $n\in\{0,\pm 1\}$, that obey the commutation relations
\begin{subequations}
\begin{align}
	\comm{J_m}{J_n}&=(m-n) J_{m+n}\,,\\
	\comm{J_m}{P_n}&=(m-n) P_{m+n}\,,\\
	\comm{P_m}{P_n}&= 0\,.
\end{align}
\end{subequations}
Generators of the algebra $\mathcal{U}(\isl(2,\R))$ can be written as formal products of these generators, modulo the respective commutation relations. In accordance with the Poincar\'{e}-Birkhoff-Witt theorem we may choose an ordering relation in which all translational generators $P_n$ are on the right-hand side and formal products on both sides are arranged in decreasing order of the mode index, such that a generic generator of $\mathcal{U}(\isl(2,\R))$ reads
\begin{align}
	\overbrace{\underbrace{J_1 \dots J_1 J_0 \dots J_0 J_{-1} \dots J_{-1}}_{l} P_1 \dots P_1 P_0\dots P_0 P_{-1}\dots P_{-1}}^{s-1}\,,
\end{align}
where we have introduced the indices $s\ge 1$ and $0\le l\le s-1$, counting the overall number of generators and the number of rotation generators, respectively. In constrast to the $\mathfrak{sl}(2,\R)$ case, here we are dealing with an infinite wedge of generators rather than an infinite tower, as depicted in figure \ref{uea_wedge}. As a third index we may introduce the sum of all mode indices, $m$, but note that the set $\{l,s,m\}$ is so far not sufficient to uniquely characterize all linearly independent generators. The product rules of $\mathcal{U}(\isl(2,\R))$ can be traced back to the equations given in Appendix~\ref{app_uea} that expand disordered formal powers of generators into linear combinations of ordered ones.

There are two quadratic Casimir elements in this algebra, namely
\begin{subequations}
\begin{align}
	\Msq&=P_0P_0-P_1P_{-1}\,,\\
	\S&=J_0P_0-\frac{1}{2}\left(J_1P_{-1}+J_{-1}P_1\right)\,,
\end{align}
\end{subequations}
that will be quotiented out in the next sub-section. On the level of formal products this can be achieved for example by formally identifying\footnote{Note that an identification of both $P_0P_0$ and $J_0P_0$ cannot consistently be implemented.} $P_0P_0\sim P_1P_{-1}+\Msq$ and $J_{-1}P_1\sim 2J_0P_0-J_1P_{-1}-2\S$\,. The result is an algebra over the field of real (or complex) numbers, equipped with an associative product. This product can still be viewed as the formal product of powers of generators, modulo commutation relations and modulo identification of Casimir elements. In the following we work out this algebra in a conveniently chosen basis.
\subsection{Highest-Weight Generators and Descendants}
We would now like to adopt the construction of algebra generators in terms of descendants of highest-weight generators. In particular, we construct generators with respect to the adjoint action\footnote{Note that the adjoint action of $J_n$ keeps the number of $P$-generators ($s-1-l$) fixed, which would not be the case for the action of $P_n$.} of $J_{-1}$; objects that commute with $J_1$ are called highest-weight generators. Thus, the strategy is to first find a complete set of such highest-weight generators and afterwards act with $\ad_{J_{-1}}=\comm{J_{-1}}{\,.\,}$ in order to define the respective descendants. The identification of Casimir elements, $P_0P_0\sim P_1P_{-1}+\Msq$ and $J_{-1}P_1\sim 2J_0P_0-J_1P_{-1}-2\S$, is done implicitly on the level of formal products.

Clearly, at fixed spin we have $s$ different combinations of maximal mode number, parame\-trized by the index $l$ (from here on, with powers of generators we denote powers of the formal product):
\begin{align}
	(J_1)^l(P_1)^{s-1-l}\,, && \text{where }\ s\ge 1\,,\ 0\le l\le s-1\,.
\end{align}
But the algebra turns out to be even larger than that. Since there is no Casimir element that eliminates higher powers of $J_0$, generators including such powers are linearly independent from the above highest-weight generators as well as their descendants. Thus, we have to include powers $(J_0)^\xi$ into the construction of highest-weight states as well.

The defining property of highest-weight generators is that they commute with $J_1$. There are two objects of order two that fulfil this requirement,
\begin{align}
	\C&\equiv (J_0)^2-J_1J_{-1}+J_0 && \text{ and } && J_0P_1-J_1P_0\,,
\end{align}
and powers of these will be the additional building blocks we need to cover all possible linearly independent formal-product expressions. The abbreviation $\C$ indicates the analogy to the Casimir element of $\sl(2,\R)$; indeed, $\C$ commutes with all generators $J_n$\,.

We propose to define highest-weight generators as
\begin{align}
	\Q{l}{s}{\xi}{s-1-\xi}:=\begin{cases}
        (J_1)^{l-\xi}\C^{\frac{\xi}{2}}(P_1)^{s-1-l}\,, & \xi \text{ even}\\[0.15cm]
        (J_1)^{l-\xi}\C^{\left\lfloor\frac{\xi}{2}\right\rfloor}\left(J_0P_1-J_1P_0\right)(P_1)^{s-2-l}\,, & \xi \text{ odd}
    \end{cases}\ \ .
\end{align}
Here $\left\lfloor \cdot\right\rfloor$ denotes the floor function. Note that, obviously, this choice is not unique. The complete set of generators is defined through repeated action of the adjoint of $J_{-1}$ on these highest-weight generators,
\begin{align}\label{def_Q}
	\Q{l}{s}{\xi}{m}:=(-1)^{s-1-\xi-m}\frac{(s+m-\xi-1)!}{(2s-2\xi-2)!}\underbrace{\Big[J_{-1},\Big[J_{-1},\dots,\Big[J_{-1}\,,}_{s-1-\xi-m}\Q{l}{s}{\xi}{s-1-\xi}\Big]\dots\Big]\Big]\,,
\end{align}
and one finds the range of indices to be
\begin{align}
	s\ge 1\,, && 0\le \xi\le 2\left\lfloor\frac{s-1}{2}\right\rfloor\,, && |m|\le s-1-\xi\,, && \xi\le l\le s-\frac{3-(-1)^\xi}{2}\,.
\end{align}
One may count the number of possible combinations of indices permitted by these ranges, which is given by $(s^2+s+1)(s+1)s/6$ up to a fixed value of $s$. This can also be derived by counting all different formal-product combinations up to order $s-1$. Keep in mind that this particular index structure is a consequence of the identification of Casimir elements; within $\mathcal{U}(\isl(2,\R))$ we would at least need another additional index to count higher powers of $P_0$.

This definition of generators naturally provides a simple behavior of the adjoint action of the element $J_n$,
\begin{align}\label{adjointJ}
	\left[\Q{l}{s}{\xi}{m}\,, J_n\right]=\left(m-(s-1-\xi)n\right)\Q{l}{s}{\xi}{m+n}\,,
\end{align}
which tells us that the mode index $m$ is indeed the eigenvalue of a generator with respect to the adjoint action of $J_0$, especially $m=s-1-\xi$ are the eigenvalues of highest-weight generators.

Due to the complicated structure of the algebra, we are at present not able to provide the full set of structure constants associated to the product (or the commutator) of two generic algebra generators. Yet, it is possible to find expressions for the left- and right-multiplication of a general $\Q{l}{s}{\xi}{m}$ with a spin-2 generator. These products are of the form
\begin{subequations}\label{left_right_s2_mult}
\begin{align}
	\Q{l}{s}{\xi}{m}\star J_n &=\sum_{\sigma=0}^3\sum_{\lambda=0}^2\sum_{\eta=0}^3 \StrGamma{\text{\scriptsize (R)}}{\lambda,\sigma}{}{\eta}(l,s,\xi;m,n) \Q{l+1-\lambda}{s+1-\sigma}{\xi+2-\eta}{m+n}\,,\\
	J_n \star \Q{l}{s}{\xi}{m} &=\sum_{\sigma=0}^3\sum_{\lambda=0}^2\sum_{\eta=0}^3 \StrGamma{\text{\scriptsize (L)}}{\lambda,\sigma}{}{\eta}(l,s,\xi;m,n) \Q{l+1-\lambda}{s+1-\sigma}{\xi+2-\eta}{m+n}\,,\\
	\Q{l}{s}{\xi}{m}\star P_n&=\sum_{\sigma=0}^3\sum_{\lambda=0}^2\sum_{\eta=0}^3 \StrGammaB{\text{\scriptsize (R)}}{\lambda,\sigma}{}{\eta}(l,s,\xi;m,n) \Q{l-\lambda}{s+1-\sigma}{\xi+2-\eta}{m+n}\,,\\
	P_n\star \Q{l}{s}{\xi}{m}&=\sum_{\sigma=0}^{\xi+2}\sum_{\lambda=0}^{\xi+2}\sum_{\eta=0}^{\xi+2} \StrGammaB{\text{\scriptsize (L)}}{\lambda,\sigma}{}{\eta}(l,s,\xi;m,n) \Q{l-\lambda}{s+1-\sigma}{\xi+2-\eta}{m+n}
\end{align}
\end{subequations}
and are explicitly written down in Appendix~\ref{app_largeUEA}.

Note that we may attach a length scale to the generators, since we may think of the spin-2 translations as possessing an inverse length dimension of one. Accordingly, a generic generator $\Q{l}{s}{\xi}{m}$ may be assigned inverse length dimension $s-1-l$. Mass Casimir $\Msq$ and spin Casimir $\S$ thus are of inverse length dimension two and one, respectively, as one may anticipate from standard quantum field theory.
\subsection{Sub-Structures of \texorpdfstring{$\ihs(\Msq,\S)$}{ihs(M\texttwosuperior,S)}}
In some of the later sections we will point out the necessity to study smaller sub-structures of $\ihs(\Msq,\S)$. In order to keep purely algebraic considerations contained within this section, we discuss two of these structures in the following.

By construction, the Lie algebra $\isl(2,\R)$ itself is contained as a Lie-subalgebra spanned by the set of generators $\{\Q{l}{2}{0}{m}\,|\, l=0,1\}$\,, where the commutator is naturally given with respect to the associative product.

Readers only interested in the relevant flat-space set-up may skip sub-section \ref{subsec_leftSlice} and proceed with \ref{subsub_rightslice}.
\subsubsection{The Left Slice}\label{subsec_leftSlice}
\begin{figure}[h]
	%\centering
	\begin{subfigure}[t]{0.45\linewidth}
	\scalebox{0.73}{
	\begin{tikzpicture}
	%\draw[fill=lightgray,color=lightgray,rounded corners,rotate=45](-0.4,-0.4) rectangle (7,0.5);
	%
	\path [fill=lightgray] (0,-0.5)--(0.5,0)--(-3.2,3.7)--(-4.2,3.7)--(0,-0.5);
	\node at(0,0){$\mathds{1}$};
	\node at(-1,1){$J$};
	\node at(1,1){$P$};
	\node at(-2,2){$JJ$};
	\node at(0,2){$JP$};
	\node at(2,2){$PP$};
	\node at(-3,3){$JJJ$};
	\node at(-1,3){$JJP$};
	\node at(1,3){$JPP$};
	\node at(3,3){$PPP$};
	\draw[-](-0.3,0.3)--(-0.7,0.7);
	\draw[-](0.3,0.3)--(0.7,0.7);
	\draw[-](-1.3,1.3)--(-1.7,1.7);
	\draw[-](1.3,1.3)--(1.7,1.7);
	\draw[-](-2.3,2.3)--(-2.7,2.7);
	\draw[-](2.3,2.3)--(2.7,2.7);
	\draw[dotted,thick](-3.3,3.3)--(-3.7,3.7);
	\draw[dotted,thick](3.3,3.3)--(3.7,3.7);
	\draw[-](-0.7,1.3)--(-0.3,1.7);
	\draw[-](0.3,2.3)--(0.7,2.7);
	\draw[dotted,thick](1.3,3.3)--(1.7,3.7);
	\draw[dotted,thick](-0.7,3.3)--(-0.3,3.7);
	\draw[-](-1.7,2.3)--(-1.3,2.7);
	\draw[dotted,thick](-2.7,3.3)--(-2.3,3.7);
	\draw[-](0.7,1.3)--(0.3,1.7);
	\draw[-](-0.3,2.3)--(-0.7,2.7);
	\draw[dotted,thick](-1.3,3.3)--(-1.7,3.7);
	\draw[dotted,thick](0.7,3.3)--(0.3,3.7);
	\draw[-](1.7,2.3)--(1.3,2.7);
	\draw[dotted,thick](2.7,3.3)--(2.3,3.7);
	\node at(-5.5,0){$s=1$};
	\node at(-5.5,1){$s=2$};
	\node at(-5.5,2){$s=3$};
	\node at(-5.5,3){$s=4$};
	%
	%\node[rotate=45] at(4.5,4.5){$l=0$};
	%\node[rotate=45] at(2.5,4.5){$l=1$};
	%\node[rotate=45] at(0.5,4.5){$l=2$};
	\node[rotate=-45] at(-4.5,4.5){$l=s-1$};
	\end{tikzpicture}
	}%
	\caption{The set of generators with index $l=s-1$ forms a sub-algebra of $\ihs(\Msq,\S)$, which we call $\ihs^{\tiny (s-1)}$.}
	\label{fig_leftslice}
	\end{subfigure}
\hspace{1.3cm}
	\begin{subfigure}[t]{0.4\linewidth}
	\scalebox{0.73}{
	\begin{tikzpicture}
	%\draw[fill=lightgray,color=lightgray,rounded corners,rotate=45](-0.4,-0.4) rectangle (7,0.5);
	%
	\path [fill=lightgray] (0,-0.5)--(1.5,1)--(-1.2,3.8)--(-4.2,3.8)--(0,-0.5);
	\node at(0,0){$\mathds{1}$};
	\node at(-1,1){$J$};
	\node at(1,1){$P$};
	\node at(-2,2){$JJ$};
	\node at(0,2){$JP$};
	\node[lightgray] at(2,2){$PP$};
	\node at(-3,3){$JJJ$};
	\node at(-1,3){$JJP$};
	\node[lightgray] at(1,3){$JPP$};
	\node[lightgray] at(3,3){$PPP$};
	\draw[-](-0.3,0.3)--(-0.7,0.7);
	\draw[-](0.3,0.3)--(0.7,0.7);
	\draw[-](-1.3,1.3)--(-1.7,1.7);
	\draw[lightgray,-](1.3,1.3)--(1.7,1.7);
	\draw[-](-2.3,2.3)--(-2.7,2.7);
	\draw[lightgray,-](2.3,2.3)--(2.7,2.7);
	\draw[dotted,thick](-3.3,3.3)--(-3.7,3.7);
	\draw[lightgray,dotted,thick](3.3,3.3)--(3.7,3.7);
	\draw[-](-0.7,1.3)--(-0.3,1.7);
	\draw[lightgray,-](0.3,2.3)--(0.7,2.7);
	\draw[lightgray,dotted,thick](1.3,3.3)--(1.7,3.7);
	\draw[lightgray,dotted,thick](-0.7,3.3)--(-0.3,3.7);
	\draw[-](-1.7,2.3)--(-1.3,2.7);
	\draw[dotted,thick](-2.7,3.3)--(-2.3,3.7);
	\draw[-](0.7,1.3)--(0.3,1.7);
	\draw[-](-0.3,2.3)--(-0.7,2.7);
	\draw[dotted,thick](-1.3,3.3)--(-1.7,3.7);
	\draw[lightgray,dotted,thick](0.7,3.3)--(0.3,3.7);
	\draw[lightgray,-](1.7,2.3)--(1.3,2.7);
	\draw[lightgray,dotted,thick](2.7,3.3)--(2.3,3.7);
	%
	%\node at(-6,0){$s=1$};
	%\node at(-6,1){$s=2$};
	%\node at(-6,2){$s=3$};
	%\node at(-6,3){$s=4$};
	%
	\node[rotate=-45] at(-4.5,4.5){$l=s-1$};
	\node[rotate=-45] at(-2.5,4.5){$l=s-2$};
	\end{tikzpicture}
	}%
	\caption{A truncation in powers of $P_n$ yields a sub-structure, only containing generators with index $l=s-1$ or $l=s-2$.}
	\label{fig_leftslicetruncated}
\end{subfigure}
\caption{On the left-hand side of the algebra, i.e. in the region of maximal $l$, we can identify two sub-structures of $\ihs(\Msq,\S)$.}
\end{figure}
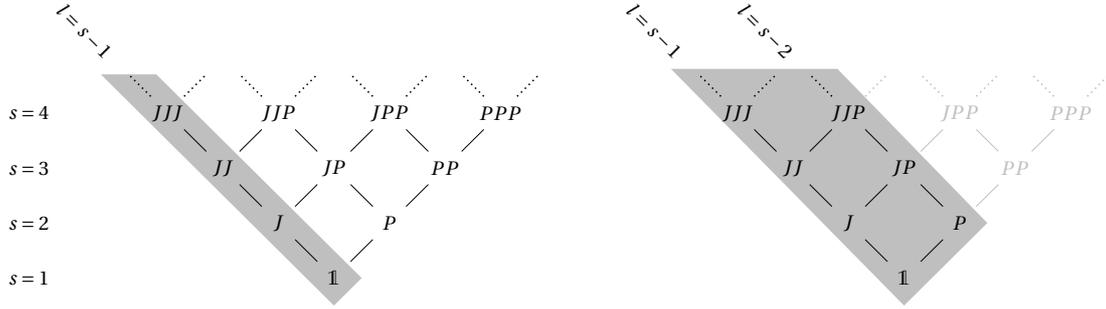

It is apparent from the construction that the set of generators with index $l=s-1$ forms the sub-algebra depicted in figure \ref{fig_leftslice} that we call $\ihs^{\tiny (s-1)}$. Within this sub-algebra all generators consist of formal powers of $J_n$\, only and thus the combination $\C\equiv (J_0)^2-J_1J_{-1}+J_0$ becomes a Casimir element of this algebra. One may then recover the AdS higher-spin algebra $\hs(\lambda)$ as the quotient algebra with respect to the ideal generated by $\C$,
\begin{align}
    \hs(\lambda)\simeq\frac{\ihs^{\tiny (s-1)}}{\langle\C\rangle}\,, && \C=\frac{\lambda^2-1}{4}\,.
\end{align}
In the language of $\ihs$-generators this quotienting amounts to the formal replacement
\begin{align}
	\Q{s-1}{s}{\xi}{m}\mapsto \C^{\frac{\xi}{2}}V^{s-\xi}_{m}\,,
\end{align}
where $\C$ is just a real (or complex) number and the generators of the quotient algebra are now called $V^s_m$ with $s\ge 1$ and $|m|\le s-1$\,. Note that $\xi$ is necessarily an even number in this case. The multiplication rules for this algebra are known as the so-called \emph{lone-star product} \cite{POPE1990191,Gaberdiel:2011wb},
\begin{align}
    V^s_m\star V^t_n&=\sum_{u=0}^{s+t-|s-t|-2} g^{st}_u (m,n;\lambda) V^{s+t-1-u}_{m+n}\,,
\end{align}
with structure constants
\begin{align}\label{lone-star_SC}
	g^{st}_u(m,n;\lambda)=\frac{1}{4^u u!}\mathcal{N}^{st}_u(m,n)\ \tensor*[_{4\!}]{F}{_3}\left[\left.\begin{matrix}	
			\sfrac{1}{2}+\lambda\,, & \sfrac{1}{2}-\lambda\,, & \sfrac{1}{2}-\sfrac{u}{2}\,, & -\sfrac{u}{2}\\
			\sfrac{3}{2}-s\,, & \sfrac{3}{2}-t\,, & \multicolumn{2}{c}{s+t-\sfrac{1}{2}-u}\hspace{1ex}
	\end{matrix}\right|\ 1\right]\,,
\end{align}
where
\begin{align}\label{lone-star_N}
	\mathcal{N}^{st}_u(m,n)=\sum_{k=0}^u (-1)^k \binom{u}{k}(s-1+m)^{\underline{u-k}}\,(s-1-m)^{\underline{k}}\,(t-1+n)^{\underline{k}}\,(t-1-n)^{\underline{u-k}}\,.
\end{align}
Here we have chosen a particular normalization already. Throughout this work we use the compact notation $a^{\underline{k}}$ and $a^{\overline{k}}$ for falling and rising factorials, respectively.

Note that one may easily reverse the quotienting in order to derive an expression for the structure constants of $\ihs^{\tiny (s-1)}$ by expanding $g^{st}_u$ into a polynomial in $\C$ and treating it as a generator again.

Let us go one step further: Imposing a formal identification of the form 
\begin{align}\label{JP_identification}
J_mP_n\sim P_mJ_n=J_nP_m+(m-n)P_{m+n}\,,
\end{align}
which may be motivated by an \.In\"on\"u-Wigner contraction (see also section \ref{sec_Inonu_Wigner}), requires a truncation in formal powers of $P_n$ in order to avoid a contradiction to the commutation relations of $\isl(2,\R)$\,. In particular, all formal powers of translational generators greater than one need to be set to zero, $P_mP_n\sim 0$ and in particular, $\Msq=0$\,.

Such a truncation can be consistently applied and it transforms the algebra $\ihs(\Msq,\S)$ into a smaller structure, only consisting of generators with indices $l=s-1$ and $l=s-2$, see figure \ref{fig_leftslicetruncated}. Furthermore, the identification (\ref{JP_identification}) gives the object $\C$ the property to commute with all generators $P_n$ and, therefore, upgrades it to a Casimir element. The distinction between generators of even and odd index $\xi$ becomes obsolete. Transcending to the quotient algebra with respect to $\C$ amounts to the formal identification
\begin{align}\label{left_slice_repl}
    \Q{l}{s}{\xi}{m}\mapsto (-1)^\xi\C^{\left\lfloor\frac{\xi}{2}\right\rfloor}\begin{cases}
        V^{s-\xi}_m\,, & s-1-l=0\\
        W^{s-\xi}_m\,, & s-1-l=1\\
        0\,, & s-1-l\ge 2
    \end{cases}\ \ ,
\end{align}
where $\C$ is now a real (or complex) number and the generators of the quotient algebra are called $V^s_m$ and $W^s_m$ with $s\ge 2$ and $|m|\le s-1$, together with the unit element $V^1_0$. Inserting the replacement (\ref{left_slice_repl}) into the product rules (\ref{left_right_s2_mult}) gives
\begin{subequations}\label{productsVJPWJP}
\begin{align}
\begin{split}\label{prodVJ}
  V^s_m\star J_n&=V^{s+1}_{m+n}+\frac{\n{1}^s(m,n)}{2}V^s_{m+n}+\frac{\n{2}^s(m,n)}{16(s-\sfrac{1}{2})^{\underline{2}}}\left(s(s-2)-4\C\right)V^{s-1}_{m+n}\,,
\end{split}\\
\begin{split}\label{prodWJ}
    W^s_m\star J_n&=W^{s+1}_{m+n}+\frac{\n{1}^s(m,n)}{2}W^s_{m+n}\\
    &\quad +\frac{\n{2}^s(m,n)}{16(s-1)^2(s-\sfrac{1}{2})^{\underline{2}}}\bigg[s(s-2)\left((s-1)^2-4\C\right)W^{s-1}_{m+n}-4\S V^{s-1}_{m+n}\bigg]\,,
\end{split}\\
\begin{split}\label{prodVP}
    V^s_m\star P_n&=W^{s+1}_{m+n}+\frac{\n{1}^s(m,n)}{2}W^s_{m+n}\\
    &\quad +\frac{\n{2}^s(m,n)}{16(s-1)(s-\sfrac{1}{2})^{\underline{2}}}\bigg[(s-2)\left(s^{\underline{2}}+4\C\right) W^{s-1}_{m+n}-4(2s-3)\S V^{s-1}_{m+n}\bigg]\,,
\end{split}\\
\begin{split}\label{prodWP}
    W^s_m\star P_n&=-\frac{(s-2)(2s-5)\n{2}^s(m,n)}{4(s-1)^2(s-\sfrac{1}{2})^{\underline{2}}}\S W^{s-1}_{m+n}\,,
\end{split}
\end{align}
\end{subequations}
where we introduced the short-hand
\begin{subequations}\label{small_n}
\begin{align}
    \n{1}^s(m,n)&=m-(s-1)n\,,\\
    \n{2}^s(m,n)&=m^2+(s-1)(2s-3)n^2-(2s-3)mn-(s-1)^2\,.
\end{align}
\end{subequations}
While (\ref{prodVJ}) displays exactly the sub-algebra $\hs(\lambda)$, there is a subtlety concerning $\S$ in the remaining equations: In order to avoid inconsistencies it needs to be set to zero, since there are no higher powers in $P_n$ and thus clearly $\S\star P_n=0$\,. This is most apparent in the product rule (\ref{prodWP}), where, for instance, the special case $s=2$ spoils the truncation. On the other hand, one might argue that we should not have done the quotienting of the Casimir element $\S$ in the first place, but rather do the truncation first. It turns out that the latter construction gives the equations (\ref{productsVJPWJP}) with the formal replacements
\begin{align}
    \S V^s_m\mapsto \C W^s_m && \text{as well as} && \S W^s_m \mapsto 0\,,
\end{align}
that practically reverse the $\S$-quotient. The product rules we are then left with are exactly those that one can find from an \.In\"on\"u-Wigner-like contraction (applied on the level of the algebra product rather than the Lie bracket) from $\hs(\lambda)\oplus\hs(\lambda)$, namely \cite{Riegler:2016hah}
\begin{subequations}\label{product_rules_oldContraction}
\begin{align}
	V^s_m\star V^t_n&=\sum_{u=0}^{s+t-|s-t|-2} g^{st}_u (m,n;\lambda) V^{s+t-1-u}_{m+n}\,,\\
	V^s_m\star W^t_n&=\sum_{u=0}^{s+t-|s-t|-2} g^{st}_u (m,n;\lambda) W^{s+t-1-u}_{m+n}\,,\\
	W^s_m\star W^t_n&=0\,,
\end{align}
\end{subequations}
with the same structure constants of $\hs(\lambda)$ defined above. Note that the consideration of the spin-$s$-spin-2 product rules (\ref{productsVJPWJP}) is actually sufficient to infer on the complete algebra structure (\ref{product_rules_oldContraction}).

We will comment more on the role of contractions that connect AdS and flat-space higher-spin algebras in section \ref{sec_Inonu_Wigner} -- for now just note that none of the flat-space Casimir elements $\S$ and $\Msq$ appears explicitly in this structure.
\subsubsection{The Right Slice}\label{subsub_rightslice}
Even though the sub-structure discussed in the previous sub-section is fully known and easy to handle, we will not be focusing on it for our later discussion of higher-spin physics in asymptotically flat spacetime. Our main reason for doing so is that these sub-structures do not allow for a consistent coupling of a scalar field to a flat-space higher-spin background. Accordingly, we will now shift the focus to the right-hand side of our wedge, i.e. to generators of index $l=0$ and $l=1$.

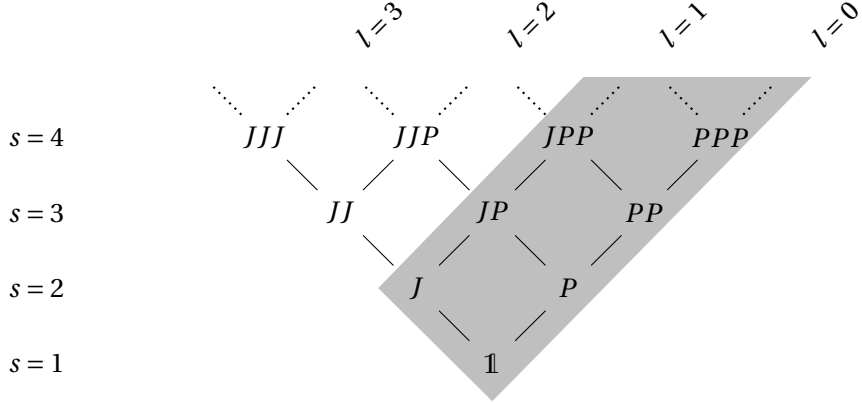
\begin{figure}[h]
    \centering
    \begin{tikzpicture}
%\draw[fill=lightgray,color=lightgray,rounded corners,rotate=45](-0.4,-0.4) rectangle (7,0.5);
%
\path [fill=lightgray] (0,-0.5)--(-1.5,1)--(1.2,3.8)--(4.2,3.8)--(0,-0.5);
\node at(0,0){$\mathds{1}$};
\node at(-1,1){$J$};
\node at(1,1){$P$};
\node at(-2,2){$JJ$};
\node at(0,2){$JP$};
\node at(2,2){$PP$};
\node at(-3,3){$JJJ$};
\node at(-1,3){$JJP$};
\node at(1,3){$JPP$};
\node at(3,3){$PPP$};
\draw[-](-0.3,0.3)--(-0.7,0.7);
\draw[-](0.3,0.3)--(0.7,0.7);
\draw[-](-1.3,1.3)--(-1.7,1.7);
\draw[-](1.3,1.3)--(1.7,1.7);
\draw[-](-2.3,2.3)--(-2.7,2.7);
\draw[-](2.3,2.3)--(2.7,2.7);
\draw[dotted,thick](-3.3,3.3)--(-3.7,3.7);
\draw[dotted,thick](3.3,3.3)--(3.7,3.7);
\draw[-](-0.7,1.3)--(-0.3,1.7);
\draw[-](0.3,2.3)--(0.7,2.7);
\draw[dotted,thick](1.3,3.3)--(1.7,3.7);
\draw[dotted,thick](-0.7,3.3)--(-0.3,3.7);
\draw[-](-1.7,2.3)--(-1.3,2.7);
\draw[dotted,thick](-2.7,3.3)--(-2.3,3.7);
\draw[-](0.7,1.3)--(0.3,1.7);
\draw[-](-0.3,2.3)--(-0.7,2.7);
\draw[dotted,thick](-1.3,3.3)--(-1.7,3.7);
\draw[dotted,thick](0.7,3.3)--(0.3,3.7);
\draw[-](1.7,2.3)--(1.3,2.7);
\draw[dotted,thick](2.7,3.3)--(2.3,3.7);
\node at(-6,0){$s=1$};
\node at(-6,1){$s=2$};
\node at(-6,2){$s=3$};
\node at(-6,3){$s=4$};
\node[rotate=45] at(4.5,4.5){$l=0$};
\node[rotate=45] at(2.5,4.5){$l=1$};
\node[rotate=45] at(0.5,4.5){$l=2$};
\node[rotate=45] at(-1.5,4.5){$l=3$};
\end{tikzpicture}
    \caption{The shaded region on the right-hand side of the algebra $\ihs(\Msq,\S)$, i.e. the set of generators with index $l=0$ or $l=1$, is what we call $\ihs^{\text{\tiny (R)}}(\Msq,\S)$. It does not constitute a sub-algebra since it does not close w.r.t. the associative product. It does, however, form a Lie-subalgebra where the Lie bracket is identified with commutators in the obvious way.}
    \label{fig_rightslice}
\end{figure}

The right slice is spanned by the set of generators $\{\Q{l}{s}{\xi}{m}\,|\, 0\le l\le 1\,, 0\le\xi\le l\,, |m|\le s-1-\xi\}$ and we will call it $\ihs^{\text{\tiny (R)}}(\Msq,\S)$. It is depicted in figure \ref{fig_rightslice}. In other words, we are dealing with descendants of the three classes of highest-weight generators
\begin{align}
    \Q{1}{s}{0}{s-1}=J_1(P_1)^{s-2}\,, && \Q{1}{s}{1}{s-2}=(J_0P_1-J_1P_0)(P_1)^{s-3}\,, && \Q{0}{s}{0}{m}=(P_1)^{s-1}\,.
\end{align}
This set does not close with respect to the associative algebra product as one may see for example by considering the product of $J_m$ and $J_n$ that produces higher-$l$ generators. Furthermore, it does not appear to be possible to find a consistent\footnote{Consistent in the sense that the commutation relations of $\isl(2,\R)$ are not violated.} truncation in formal powers of the generators $J_n$. 

However, we notice that the commutators with respect to the associative product do form a closed structure, thus providing us with a Lie algebra on which we can introduce higher-spin gauge fields. We will later see that the special form of these gauge fields leads to equations of motion that are closed in a certain sense as well.

The product rules and commutation relations of the right slice are explicitly presented in Appendix~\ref{app_strCons_right}. For now we only present their overall form, which is
\begin{subequations}\label{right_slice_multipl}
\begin{align}
	\Q{0}{s}{0}{m}\star\Q{0}{t}{0}{n}&=\sum_{u} h^{st}_u(m,n)\ \Q{0}{s+t-1-2u}{0}{m+n}\,,\\
	\begin{split}\label{middle_prod}
    \Q{1}{s}{0}{m}\star\Q{0}{t}{0}{n}&=\sum_{u}\sum_{l=0}^1\sum_{\xi=0}^l f^{st}_{u,l,\xi}(m,n)\Q{l}{s+t-1-2u}{\xi}{m+n}\,,
    \end{split}\\
    \begin{split}
    \Q{1}{s}{1}{m}\star\Q{0}{t}{0}{n}&=\sum_{u}\sum_{l=0}^1\sum_{\xi=0}^l \bar{f}^{st}_{u,l,\xi}(m,n)\Q{l}{s+t-1-2u}{\xi}{m+n}\,,
    \end{split}\\[0.15cm]
    \comm{\Q{1}{s}{0}{m}}{\Q{0}{t}{0}{n}}&=\sum_u d^{st}_u(m,n) \Q{0}{s+t-2-2u}{0}{m+n}\,,\\
    \comm{\Q{1}{s}{1}{m}}{\Q{0}{t}{0}{n}}&=\sum_u \bar{d}^{st}_u(m,n) \Q{0}{s+t-2-2u}{0}{m+n}\,,\\
    \comm{\Q{1}{s}{0}{m}}{\Q{1}{t}{0}{n}}&=\sum_u\sum_{l=0}^1 b^{st}_{u,l}(m,n) \Q{l}{s+t-2-2u}{0}{m+n}\,.
\end{align}
\end{subequations}
We did not include the commutation relations $\comm{\Q{1}{s}{1}{m}}{\Q{1}{t}{0}{n}}$ and $\comm{\Q{1}{s}{1}{m}}{\Q{1}{t}{1}{n}}$, because they are neither needed for this work, nor of an overwhelmingly simple form. Note that the generators of index $\xi=0$ form a Lie subalgebra. It is precisely this subalgebra that we propose to be the appropriate Lie algebra for a flat-space higher-spin Chern-Simons theory.

As a first step, one may convince oneself that the $\isl(3,\R)$-proposal of \cite{Afshar:2013vka} is contained within $\ihs^{\text{\tiny (R)}}(\Msq,\S)$. To see this, let us perform a truncation to spin 3, i.e. identify all generators of spin greater than 3 with zero. Call $U_m=\Q{1}{3}{0}{m}$ and $V_m=\frac{1}{2}\Q{0}{3}{0}{m}$, as well as $R_m\equiv\Q{1}{3}{1}{m}$. This yields the following Lie algebra:
\begin{subequations}\label{isl3}
\begin{align}
    [J_m,J_n]&=(m-n)J_{m+n}\,, &\hspace{0.6cm} [J_m,P_n]&=(m-n)P_{m+n}\,,\hspace{1.2cm} [P_m,P_n]=0\,,\\[0.5cm]
%\end{align}
%\begin{align}
    [U_m,J_n]&=(m-2n)U_{m+n}\,,     & [U_m,P_n]&=(m-2n)V_{m+n}  \vphantom{\frac{2\Msq}{3}}\,,\\
    [V_m,J_n]&=(m-2n)V_{m+n}\,,     & [V_m,P_n]&=0\,,  \vphantom{\frac{2\Msq}{3}}\\
    [R_m,J_n]&=(m-n)R_{m+n}\,,      & [R_m,P_n]&=-2V_{m+n}-\frac{2\Msq}{3}(m^2+n^2-mn-1)\mathds{1}\,,
\end{align}
\begin{align}
    [U_m,U_n]&=-\frac{(m-n)(2m^2+2n^2-mn-8)}{60}\left(\Msq J_{m+n}+2\S P_{m+n}\right)\,,\\
    [V_m,V_n]&=0\,,\\
    [R_m,R_n]&=\Msq(m-n)J_{m+n}\,,\\
    [U_m,V_n]&=-\frac{\Msq(m-n)(2m^2+2n^2-mn-8)}{60}P_{m+n}\,,\\
    [U_m,R_n]&=\frac{m^2+6n^2-3mn-4}{5}\left(\Msq J_{m+n}-\frac{\S}{2}P_{m+n}\right)\,,\\
    [V_m,R_n]&=\frac{\Msq(m^2+6n^2-3mn-4)}{10}P_{m+n}\,.
\end{align}
\end{subequations}
Apparently, one needs to consider the sub-algebra spanned by generators $J_m$, $P_m$, $U_m$, $V_m$ and set $\S=0$ as well as $\Msq=-60$ to arrive at the commutation relations of \cite{Afshar:2013vka}.
\section{The Higher-Spin \texorpdfstring{\.In\"on\"u}{Inonu}-Wigner Contraction Revisited}\label{sec_Inonu_Wigner}
There are many known cases of different Lie algebras that are connected by \.In\"on\"u-Wigner contractions \cite{Inonu510}. For our purposes we are interested in a contraction from higher-spin algebras that are used in an AdS context to higher-spin algebras that can be applied in a flat-space context. The physical intuition underlying such a contraction is that one should encounter a flat-space scenario when taking the AdS radius to infinity\footnote{Of course, the situation is not quite so simple in general. In fact, the interactions between massless higher-spin fields are not analytic in the cosmological constant \cite{FRADKIN198789}.} or equivalently the cosmological constant to zero.

In this section we argue how it should be possible to extend such contractions from the level of Lie brackets to the level of associative products. The guiding principle of our considerations is the compatibility of the formal-product constructions on both the AdS and the flat-space side. 

In order to make our arguments more clear and to provide an explicit, tractable example we make use of a non-relativistic contraction of the Lie algebra $\isl(2,\R)$ from two copies of $\sl(2,\R)$ and show how to obtain contraction rules that are compatible with the UEA construction. It is important to note that the same arguments that apply to the non-relativistic construction that we present here can also be applied to an ultra-relativistic contraction. However, the resulting expressions turn out to be much more complicated and thus for the sake of simplicity we decide to focus on the non-relativistic contraction in this section.
\subsection{Contraction from a Formal-Product Point of View}
When we consider the contraction on the classical level, $\sl(2,\R)\oplus\sl(2,\R)\rightarrow\isl(2,\R)$, the procedure is to linearly combine generators $L_n\in\sl(2,\R)$ and $\bar{L}_n\in\overline{\sl}(2,\R)$ according to\footnote{The sign in $P_n$ is a matter of convention -- it has no impact on the resulting Lie algebra.}
\begin{subequations}\label{contraction_spin2}
\begin{align}
	J_n&=L_n+\bar{L}_n\,,\\
	P_n&=-\epsilon\left(L_n-\bar{L}_n\right)\,,
\end{align}
\end{subequations}
compute the commutators of $J_n$ and $P_n$, and eventually take the limit $\epsilon\rightarrow 0$\,. Intuitively, this principle may be continued to the higher-spin case by combining the generators $\Vsc^s_m$ and $\Vb^s_m$ of the AdS (Lie) algebra $\hs(\lambda)\oplus\hs(\lambda)$ like
\begin{subequations}\label{contraction_old}
\begin{align}
	V^s_m&=\Vsc^s_m+\Vb^s_m\,,\\
	W^s_m&=-\epsilon\left(\Vsc^s_m-\Vb^s_m\right)\,.
\end{align}
\end{subequations}
This approach yields the product rules (\ref{product_rules_oldContraction}) when applied with respect to the associative product of $\hs(\lambda)\oplus\hs(\lambda)$ and the respective Lie structure constants when applied to the Lie bracket.

Clearly, the procedure does not yield the complete flat-space higher-spin algebra $\ihs(\Msq,\S)$ (which is obvious already from the different sizes) but at most some sub-structure of the latter, which we cannot be certain to be a physically significant object. Indeed, we have already constructed that object in sub-section \ref{subsec_leftSlice}, where we could see that it actually emerges from a truncation of $\mathcal{U}(\isl(2,\R))$\,, in particular setting quadratic and higher powers of $P_n$ to zero\footnote{That this truncation is induced by the contraction is most evident when taking the Grassmann path to flat space \cite{Krishnan:2013wta}; then translational generators are proportional to some Grassmann variable that squares to zero, $P_n\sim\chi$ with $\chi^2=0$\,.}, and subsequently taking the quotient with respect to the object $\C=(J_0)^2-J_1J_{-1}+J_0$\,, which became a Casimir element due to the truncation. One may easily see this connection by replacing the respective highest-weight generators by AdS generators,
\begin{subequations}
\begin{align}
    V^s_{s-1}&=(J_1)^{s-1}=\left(L_1+\bar{L}_1\right)^{s-1}\sim (L_1)^{s-1}+(\bar{L}_1)^{s-1}=\Vsc^{s}_{s-1}+\Vb^s_{s-1}\,,\\
    \begin{split}
    W^s_{s-1}&=(J_1)^{s-2}P_1=-\epsilon\left(L_1+\bar{L}_1\right)^{s-2}\left(L_1-\bar{L}_{1}\right)\sim -\epsilon\left((L_1)^{s-1}-(\bar{L}_1)^{s-1}\right)\\
    &=-\epsilon\left(\Vsc^s_{s-1}-\Vb^s_{s-1}\right)\,.
    \end{split}
\end{align}
\end{subequations}
It is crucial that we work in $\hs(\lambda)\oplus\hs(\lambda)$, implying $L_m\bar{L}_n\sim 0$\,. Keep in mind that there are no flat-space Casimir elements around anymore: from the viewpoint of the quotient of the truncated UEA $\Msq$ never existed and $\S$ just was not divided out.

From the point of view of a contraction, the following happens: When imposing the (non-relativistic) contraction naively, one would define new Casimir elements through
\begin{subequations}
\begin{align}
    \tilde{\mathcal{M}}^2&=\C_{\text{\tiny AdS}}+\bar{\C}_{\text{\tiny AdS}}\,,\\
    \S&=-\epsilon\left(\C_{\text{\tiny AdS}}-\bar{\C}_{\text{\tiny AdS}}\right)\,.
\end{align}
\end{subequations}
The second line matches the definition $\S=J_0P_0-\frac{1}{2}\left(J_1P_{-1}+J_{-1}P_1\right)$, but in order to achieve a contraction of the product rules without divergent contributions of $1/\epsilon$, one has to enforce $\C_{\text{\tiny AdS}}=\bar{\C}_{\text{\tiny AdS}}$ in the first place, such that $\S$ necessarily vanishes. Furthermore, when comparing with the flat-space mass Casimir one finds that it does not match the expression in the first line! Instead we find $\tilde{\mathcal{M}}^2=\C$, which became a Casimir element only after truncation.

In order to access the whole of $\ihs(\Msq,\S)$, a contraction must be set up in a way compatible with the formal-product construction. Setting up an algebra contraction amounts to replacing the fundamental building blocks of $\ihs(\Msq,\S)$, i.e. the spin-2 generators $J_m$ and $P_m$, by the linear combinations (\ref{contraction_spin2}). This naturally spoils the simple contraction rule (\ref{contraction_old}) since it produces products of the form $L_m\bar{L}_n$\,, which cannot be set to zero without forcing a truncation of $\ihs(\Msq,\S)$\,. In other words: The starting point of a contraction to $\ihs(\Msq,\S)$ necessarily has to be an algebra of the same size. This algebra is a larger version of the AdS higher-spin algebra that also contains mixed products of barred and unbarred generators.

Note further that the contraction parameter $\epsilon$ can be thought of as the inverse radius of curvature of AdS, so being equipped with a length scale. This is in accordance with our remark about the length scale of $\ihs$-generators in the previous section.
\subsection{The Large AdS Higher-Spin Algebra \texorpdfstring{$\hs^{(+)}(\lambda,\bar{\lambda})$}{hs(+)(lambda,lambda-bar)}}
In the previous section we convinced ourselves that a contraction from an AdS higher-spin algebra to a flat-space higher-spin algebra that is compatible with the formal-product point of view on both sides, i.e. that admits a UEA-construction for both higher-spin algebras, necessarily needs as a starting point an AdS algebra that is of the same size as the flat-space one, namely that contains mixed products $L_m\bar{L}_n$\,. To explore this connection we have to appreciate the difference between the associative algebras $\mathcal{U}(\sl(2,\R))\oplus\mathcal{U}(\sl(2,\R))$ and $\mathcal{U}(\sl(2,\R)\oplus\sl(2,\R))$\,.

Let us repeat the analogous construction of section \ref{sec_ihsConstruction} for the case of $\sl(2,\R)\oplus\sl(2,\R)$. We will call the associative algebra structure emerging from that construction
\begin{align}
\hs^{(+)}(\lambda,\bar{\lambda})=\frac{\mathcal{U}(\sl(2,\R)\oplus\sl(2,\R))}{\left\langle \C_{\text{\tiny AdS}}\,,\bar{\C}_{\text{\tiny AdS}}\right\rangle}\,,
\end{align}
where $\C_{\text{\tiny AdS}}=(L_0)^2-L_1L_{-1}+L_0$ and $\bar{\C}_{\text{\tiny AdS}}=(\bar{L}_0)^2-\bar{L}_1\bar{L}_{-1}+\bar{L}_0$ denote the quadratic Casimir elements of the two copies of $\sl(2,\R)$\,, parametrized by $\lambda$ and $\bar{\lambda}$ in the usual way.

The spin-2 generators fulfil the commutation relations
\begin{subequations}
\begin{align}
	[L_m\,,L_n]&=(m-n)L_{m+n}\,,\\
	[\bar{L}_m\,,\bar{L}_n]&=(m-n)\bar{L}_{m+n}\,,\\
	[L_m\,,\bar{L}_n]&=0\,.
\end{align}
\end{subequations}
We may construct all generators as descendants of suitable highest-weight generators with respect to the adjoint action of $L_n+\bar{L}_{n}$\,. There are two second-order objects that commute with $L_1+\bar{L}_1$\,, namely
\begin{align}
    \mathcal{D}\equiv 2L_0\bar{L}_0-L_1\bar{L}_{-1}-L_{-1}\bar{L}_1 && \text{as well as} && L_0\bar{L}_1-L_1\bar{L}_0\,.
\end{align}
Thus, we may define highest-weight generators as\footnote{We again distinguish the cases $\xi$ being even or odd, as it was the case for $\ihs(\Msq,\S)$\,. Here we just chose a compact notation.}
\begin{align}
    \V{l}{s}{\xi}{s-1-\xi}:=(L_1)^{l-\left\lfloor\frac{\xi+1}{2}\right\rfloor}\mathcal{D}^{\left\lfloor\frac{\xi}{2}\right\rfloor}(L_0\bar{L}_1-L_1\bar{L}_0)^{\left\lfloor\frac{\xi+1}{2}\right\rfloor-\left\lfloor\frac{\xi}{2}\right\rfloor}(\bar{L}_1)^{s-1-l-\left\lfloor\frac{\xi+1}{2}\right\rfloor}\,.
\end{align}
The complete set of generators is then defined through
\begin{align}
    \V{l}{s}{\xi}{m}:=(-1)^{s-1-\xi-m}\frac{(s+m-\xi-1)!}{(2s-2\xi-2)!}\underbrace{\Big[L_{-1}+\bar{L}_{-1},\Big[ L_{-1}+\bar{L}_{-1},\dots,\Big[ L_{-1}+\bar{L}_{-1},}_{s-1-\xi-m}\V{l}{s}{\xi}{s-1-\xi}\Big]\dots\Big]\Big]
\end{align}
and the range of indices is
\begin{align}
    s\ge 1\,, && 0\le\xi\le 2\left\lfloor\frac{s-1}{2}\right\rfloor\,, && |m| \le s-1-\xi\,, && \left\lfloor\frac{\xi+1}{2}\right\rfloor\le l \le s-1-\left\lfloor\frac{\xi+1}{2}\right\rfloor\,,
\end{align}
which can be checked to give the same number of generators (per spin) as the flat-space algebra $\ihs(\Msq,\S)$. Note that $\mathcal{D}$ actually commutes with $L_n+\bar{L}_n$ for all $n$. One may convince oneself that
\begin{align}
    \left[\V{l}{s}{\xi}{m}\,,L_n+\bar{L}_n\right]=\left(m-(s-1-\xi)n\right)\V{l}{s}{\xi}{m+n}\,.
\end{align}

The standard higher-spin algebra $\hs(\lambda)\oplus\hs(\lambda)$ naturally emerges as the direct sum of the sub-algebras spanned by $(l=s-1)$- and $(l=0)$-generators, $\Vsc^s_m=\V{s-1}{s}{0}{m}$ and $\Vb^s_m=\V{0}{s}{0}{m}$. Equivalently, this can be viewed as a truncation, i.e. setting products of barred and unbarred spin-2 generators to zero\footnote{Note that, in order to avoid inconsistencies with the Casimir elements, one then needs to formally assign them two different unit elements, a barred and an unbarred one.}.

\begin{figure}[h]
\begin{center}
\begin{tikzpicture}
\node at(0,0){$\hs(\lambda)$};%$\frac{\mathcal{U}(\sl)}{\langle\C_2\rangle}\oplus\frac{\mathcal{U}(\sl)}{\langle\bar{\C}_2\rangle}$};
\node at(6.1,0){$\hs^{(+)}(\lambda,\bar{\lambda})$};
\node at(12.3,0){$\ihs(\Msq,\S)$};
\draw[{Latex[width=4pt, length=8pt]}-,thick](1.15,0)--(4.5,0);
\draw[-{Latex[width=4pt, length=8pt]},thick](7.7,0)--(11,0);
\node[above]at(3.1,0){\scriptsize sub-algebra};
\node[above]at(9.3,0){\scriptsize contraction};
\draw[{Latex[width=4pt, length=8pt]}-{Latex[width=4pt, length=8pt]},thick](0,-0.4)--(0,-1.6);
\draw[{Latex[width=4pt, length=8pt]}-{Latex[width=4pt, length=8pt]},thick](6.1,-0.4)--(6.1,-1.6);
\draw[{Latex[width=4pt, length=8pt]}-{Latex[width=4pt, length=8pt]},thick](12.3,-0.4)--(12.3,-1.6);
\node at(0,-2){$\sl(2,\R)$};
\node at(6.1,-2){$\sl(2,\R)\oplus\sl(2,\R)$};
\node at(12.3,-2){$\isl(2,\R)$};
\draw[{Latex[width=4pt, length=8pt]}-,thick](1.15,-2)--(4.5,-2);
\draw[-{Latex[width=4pt, length=8pt]},thick](7.7,-2)--(11,-2);
\node[above]at(3.1,-2){\scriptsize sub-algebra};
\node[above]at(9.3,-2){\scriptsize contraction};
\end{tikzpicture}
\end{center}
\caption{The path from AdS to flat space. The full algebra $\ihs(\Msq,\S)$ may only be obtained as a contraction from AdS when taking a detour through the larger algebra $\hs^{(+)}(\lambda,\bar{\lambda})$, which is a quotient of $\mathcal{U}(\sl(2,\R)\oplus\sl(2,\R))$.}\label{fig_path_AdS_flat}
\end{figure}
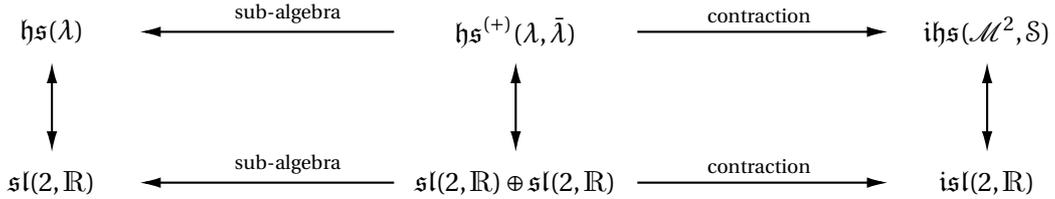

It is not within the realm of this paper to present the multiplication rules of the algebra $\hs^{(+)}(\lambda,\bar{\lambda})$. The point we want to make is that the path from higher-spin AdS to higher-spin flat space should better be viewed on the level of the larger algebras $\hs^{(+)}(\lambda\,,\bar{\lambda})$ and $\ihs(\Msq,\S)$, even if the physically relevant objects might be sub-structures thereof. This picture is partly illustrated in figure \ref{fig_path_AdS_flat}.

The easiest example of how the connection between the AdS generators $\V{l}{s}{\xi}{m}$ and the flat-space generators $\Q{l}{s}{\xi}{m}$ looks like, is the case $\xi=0$\,, for which we find
\begin{align}\label{contraction}
	\Q{l}{s}{0}{m}&=\epsilon^{s-1-l}\sum_{k=0}^{s-1}\sum_{i+j=k}(-1)^j\binom{l}{i}\binom{s-1-l}{j} \V{k}{s}{0}{m}\,.
\end{align}
Note that the description of generators in terms of a contraction is not unique since neither the definition of the highest-weight generators $\Q{l}{s}{\xi}{s-1-\xi}$ nor the definition of the highest-weight generators $\V{l}{s}{\xi}{s-1-\xi}$ are unique. The important point is the appearance of different powers of $\epsilon$, which is a fundamental difference to the simple contraction in (\ref{contraction_old}).

We considered the non-relativistic limit in this section, only. Had we used the ultra-relativistic contraction rule for the generators $J_n$ and $P_n$, the identification (\ref{contraction}) would become much more complicated -- indeed, we would need knowledge about the complete set of multiplication rules in $\hs^{(+)}(\lambda,\bar{\lambda})$, which is beyond the scope of this paper.
\section{Linear Coupling to Matter Fields}\label{sec_Vasi}
In order to study a non-trivial example of a flat-space higher-spin theory our aim is to describe the linear coupling of a scalar field to a given asymptotically flat higher-spin background. The goal is to use such a description to derive a flat-space analogue of the generalized Klein-Gordon equation presented in section \ref{sec_review}.
\subsection{Chern-Simons Formulation of Flat-Space Higher-Spin Gravity}
The gauge sector of three-dimensional higher-spin theory can be described by a non-abelian Chern-Simons theory \cite{Prokushkin:1998bq,Henneaux:2010xg,Campoleoni:2010zq}, as we have already pointed out in section \ref{sec_review}. It is well known that both asymptotically (A)dS as well as asymptotically flat spacetimes may be described by a Lie-algebra valued gauge field $A$, taking values in the algebra of isometries $\mathfrak{g}$ of the respective spacetime.

In the classical spin-2 case a large class of (2+1)-dimensional, asymptotically flat spacetimes is given by the $\isl(2,\R)$-valued gauge field \cite{Afshar:2013vka}
\begin{align}\label{flat_s2_gf}
	A=\left(P_1-\frac{M(\phi)}{4}P_{-1}\right)d\!u+\frac{1}{2}P_{-1}d\!r+\left(J_1-\frac{M(\phi)}{4}J_{-1}+r P_0-\frac{N(u,\phi)}{2}P_{-1}\right)d\!\phi\,,
\end{align}
where $(u,r,\phi)$ denote (outgoing) Eddington-Finkelstein coordinates. In a metric formulation this gauge field corresponds to
\begin{align}\label{EF_metric}
    d\!s^2=M(\phi) d\!u^2-2d\!ud\!r+2N(u,\phi) d\!ud\!\phi+r^2d\!\phi^2\,.
\end{align}
The expressions $M(\phi)$ and $N(u,\phi)$ are arbitrary functions, subject to the integrability condition $\partial_{\!\phi}M(\phi)=2\partial_{\!u}N(u,\phi)$. Different choices of these functions correspond to different spacetime geometries, e.g. $M(\phi)=-1$ and $N(u,\phi)=0$ corresponds to global Minkowski spacetime.

The gauge field may be decomposed into vielbein and spin-connection,
\begin{align}
	A=e+\omega\,,
\end{align}
where $e\sim P_n$ contains translational generators whereas $\omega\sim J_n$ contains rotational ones. In the language of vielbein and spin connection, the flatness condition $\d A+A\wedge A=0$ is equivalent to
\begin{subequations}\label{curvature_torsion}
\begin{align}
	0&=\d\omega+\omega\wedge\omega+e\wedge e\,,\\
    0&=\d e+\omega\wedge e + e\wedge\omega\,,
\end{align}
\end{subequations}
implying vanishing curvature and vanishing torsion, respectively. In the following, we will use the formulation in terms of $\omega$ and $e$ rather than $A$.

The description of a higher-spin theory of gravity requires some underlying higher-spin Lie algebra, in which the fields $e$ and $\omega$ take values in. For the case of finite higher spins one would like to extend the symmetry algebra, in our case from $\isl(2,\R)$ to $\isl(N,\R)$, and add respective higher-spin charges to the gauge field (\ref{flat_s2_gf}). For unbounded spin, $N\rightarrow\infty$, one should consider an algebra like Lie-$\ihs$ that is given by the UEA of the isometries of the corresponding metric. In turn, the finite-spin algebras $\isl(N,\R)$ can be defined as truncated versions of the latter. An example ($N=3$) is given by the authors of \cite{Afshar:2013vka}, who proposed a particular spin-3 symmetry algebra that, as shown in sub-section \ref{subsub_rightslice}, can also be obtained as truncation of (the right slice of) $\ihs(\Msq,\S)$, where the Lie bracket is identified with the commutator of the associative product in a standard way.

We propose to take the Lie-subalgebra that is spanned by generators $\Q{0}{s}{0}{m}$ and $\Q{1}{s}{0}{m}$ as flat-space higher-spin Lie algebra. Generally, we are focusing on the right slice because it contains both the flat-space Casimir elements $\Msq$ and $\S$. We drop all generators of index $l>1$ as well as the $(\xi=1)$-generators, reason being the examination of the spin-3 case. One may take the truncated algebra given in (\ref{isl3}) and add the higher-$l$ generators $\Q{2}{3}{0}{m}$ as well as $\Q{2}{3}{2}{0}$. By attempting to write down a gauge field deformed by spin-3 charges, one notices that these additional contributions are not allowed by the flatness condition. Consequently, we consider the algebra
\begin{subequations}\label{right_slice_comm}
\begin{align}
    \comm{\Q{1}{s}{0}{m}}{\Q{1}{t}{0}{n}}&=\sum_u\sum_{l=0}^1 b^{st}_{u,l}(m,n) \Q{l}{s+t-2-2u}{0}{m+n}\,,\\
    \comm{\Q{1}{s}{0}{m}}{\Q{0}{t}{0}{n}}&=\sum_u d^{st}_u(m,n) \Q{0}{s+t-2-2u}{0}{m+n}\,,\\
	\comm{\Q{0}{s}{0}{m}}{\Q{0}{t}{0}{n}}&=0
\end{align}
\end{subequations}
to be the appropriate Lie algebra, which we call Lie-$\ihs$. The explicit form of the structure constants can be found in Appendix~\ref{app_strCons_right}. 

Note that it is in principle possible to define this Lie algebra as a contraction from the AdS higher-spin algebra $\hs^{(+)}(\lambda,\bar{\lambda})$. The definition of Lie-$\ihs$ generators in terms of AdS generators is given in (\ref{contraction}). However, one needs to know the commutation rules of $\hs^{(+)}(\lambda,\bar{\lambda})$, first, to explicitly show the emergence of the structure constants in (\ref{right_slice_comm}), which is something we are not able to provide within the scope of this work.

Based on all the previous arguments, we consider now higher-spin deformations of (\ref{flat_s2_gf}) by lowest-weight generators of the form
\begin{subequations}\label{general_hs_deform}
\begin{align}
    \omega&=\omega^{\text{\tiny (class)}}+\sum_{s=3}^{\infty} W^{(s)}_\mu(u,r,\phi) d\!x^\mu \Q{1}{s}{0}{-(s-1)}\,,\\
    e&=e^{\text{\tiny (class)}}+\sum_{s=3}^{\infty} Z^{(s)}_\mu(u,r,\phi) d\!x^\mu \Q{0}{s}{0}{-(s-1)}\,,
\end{align}
\end{subequations}
thereby interpreting $(l=0)$-generators as higher-spin translations and $(l=1)$-generators as higher-spin rotations. The fact that the latter appear only in the spin connection will be of importance for a consistent coupling to matter fields later on.
\subsection{Linearized Vasiliev Gravity and Flat-Space Proposal}
We reviewed the linearization of Vasiliev's higher-spin theory in sub-section \ref{subsec_rev_linearisation} and will in the following propose to apply a variation thereof in a flat-space scenario.

Recall the equations for the geometric sector as well as for the coupling of matter to gravity:
\begin{subequations}
\begin{align}
	\d W&=W\wedge_{\!\star} W\,	\label{W_flatness}\,,\\
	\d\C&=\comm{W}{\C}_{\!\star}\,.  \label{matter_grav}
\end{align}
\end{subequations}
In close analogy to the AdS case, expressing $W$ by the gauge fields of the Chern-Simons formulation,
\begin{align}
	W=-(\omega+e)\,,
\end{align}
equation (\ref{W_flatness}), indeed, reduces to the zero-curvature and zero-torsion conditions (\ref{curvature_torsion}). Furthermore, we decompose the matter field into an auxiliary and a dynamic part,
\begin{align}
	\C=C^{\text{\tiny (aux)}}+C \psi\,,
\end{align}
where the auxiliary parameter $\psi$ is supposed to anti-commute with the algebra generators contained in the zuvielbein, $\{\psi\,,e\}=0$\,. Equation (\ref{matter_grav}) then implies
\begin{subequations}
\begin{align}
	0&=\d C^{\text{\tiny (aux)}}+\comm{\omega}{C^{\text{\tiny (aux)}}}_{\!\star}+\comm{e}{C^{\text{\tiny (aux)}}}_{\!\star}\,,\label{Vasiliev_eom_aux} \\
	0&=\d C+\comm{\omega}{C}_{\!\star}+\acomm{e}{C}_{\!\star}\,. \label{Vasiliev_eom_dyn}
\end{align}
\end{subequations}
As before, it may be argued \cite{Vasiliev:1999ba} that equation (\ref{Vasiliev_eom_aux}) describes topological fields carrying no degrees of freedom whatsoever and that $C^{\text{\tiny (aux)}}$ can thus be consistently set to zero. 

We propose (\ref{Vasiliev_eom_dyn}) as the appropriate equation to describe the propagation of a massive scalar field $C$ on the background geometry encoded in the gauge fields $\omega$ and $e$. Furthermore, the underlying star product that is required by the appearance of the anti-commutator can simply be identified with the associative algebra product emerging from the UEA-construction of the previous sections i.e. with the algebra product of $\ihs(\Msq,\S)$. This is in close analogy to the AdS case where the Moyal product is replaced by the lone-star product.

Note that we encounter a remarkable difference to the AdS case: the associative algebra structure we need to consider in the description of a matter-gravity coupling is much larger (contains a larger set of generators) than the Lie algebra structure governing the gauge sector alone.
\section{Scalar Fields in Flat-Space Higher-Spin Gravity}\label{sec_dynamics}
One may take the classical field (\ref{flat_s2_gf}) and evaluate equation (\ref{Vasiliev_eom_dyn}) by expanding the master field into \emph{all} generators of $\ihs(\Msq,\S)$,
\begin{align}
    C=\sum_{(\ihs)} \c{l}{s}{\xi}{m}(u,r,\phi)\Q{l}{s}{\xi}{m}\,,
\end{align}
where the sum is supposed to run over all allowed index combinations, while making use of the product rules (\ref{left_right_s2_mult}). It then turns out that no classical Klein-Gordon equation of the form (\ref{rev_class_KG}) can be derived. At first glance appearing like a contradiction to our UEA approach, this actually may be viewed as an indication that the complete higher-spin algebra $\ihs(\Msq,\S)$ is too large and we should rather consider a smaller sub-structure. Indeed, as we convinced ourselves in section \ref{sec_Inonu_Wigner}, a similar reduction was implicitly imposed in the case of asymptotically AdS spacetimes, where one usually considers $\hs(\lambda)\oplus\hs(\lambda)$ instead of $\hs^{(+)}(\lambda,\bar{\lambda})$.

Since we already convinced ourselves that higher-spin gauge fields are living in the Lie-subalgebra of generators $\Q{0}{s}{0}{m}$ and $\Q{1}{s}{0}{m}$, we want to find an associative structure that includes this piece but is not the complete $\ihs(\Msq,\S)$. Due to the product rules, we are forced to include generators of index $\xi=1$, therefore arriving at what we referred to as the right slice $\ihs^{\text{\tiny (R)}}(\Msq,\S)$ in sub-section \ref{subsub_rightslice}. Despite the fact that this structure does not close with respect to the algebra multiplication, we can still take it as a minimal example because it is actually sufficient to let it close with respect to the equations of motion! In other words, we simply restrict the generators $\Q{1}{s}{\xi}{m}$ to appear in the spin connection only such that they will never occur in any bilinear form other than the commutator, as is clear from the structure of the equation of motion (\ref{Vasiliev_eom_dyn}). This has already been anticipated in the general deformation (\ref{general_hs_deform}).

The reasoning behind studying the right-hand side of $\ihs(\Msq,\S)$ as an underlying algebra structure is that we clearly need to use a part of it that actually contains the mass Casimir element $\Msq$, since this is supposed to show up in connection with the mass of a scalar field propagating on the background geometry -- a feature that is missing in the left slice.
\subsection{Classical Gravity Recovered}
We will now use the right slice $\ihs^{\text{\tiny (R)}}(\Msq,\S)$ from sub-section \ref{subsub_rightslice} to evaluate the equation of motion
\begin{align}
	0&=\d C+\comm{\omega}{C}_{\!\star}+\acomm{e}{C}_{\!\star}\,,
\end{align}
where $C\equiv C(u,r,\phi)$\,. The asymptotically flat $\isl(2,\R)$-valued gauge field as given in (\ref{flat_s2_gf}) in terms of vielbein and spin connection reads
\begin{subequations}
\begin{align}
	\omega &=\left(J_1-\frac{M(\phi)}{4}J_{-1}\right)d\!\phi\,,\\
    e&=\left(P_1-\frac{M(\phi)}{4}P_{-1}\right)d\!u+\frac{1}{2}P_{-1}d\!r+\left(r P_0-\frac{N(u,\phi)}{2}P_{-1}\right)d\!\phi\,.
\end{align}
\end{subequations}
The gauge flatness condition implies the integrability condition
\begin{align}
    \partial_{\!\phi}M(\phi)=2\partial_{\!u}N(u,\phi)\,.
\end{align}
The matter field is an element of $\ihs^{\text{\tiny (R)}}(\Msq,\S)$ and therefore may be expanded in terms of generators as
\begin{align}
	C=c(u,r,\phi)\mathds{1}+\sum_{s=2}^{\infty}\ \sum_{l=0}^{1}\ \sum_{\xi=0}^{l}\ \sum_{|m|\le s-1-\xi}\  \c{l}{s}{\xi}{m}(u,r,\phi)\ \Q{l}{s}{\xi}{m}
\end{align}
and we need to apply the (anti-)commutation rules obtained from the multiplication rules (\ref{right_slice_multipl}) or (\ref{mult_Q_spin2}),
\begin{subequations}
\begin{align}
	\comm{\Q{l}{s}{\xi}{m}}{J_n}&=\left(m-(s-1-\xi)n\right)\Q{l}{s}{\xi}{m+n}\,,\\
	\begin{split}\label{middle_acomm}
	\acomm{\Q{l}{s}{0}{m}}{P_n}&=2\Q{l}{s+1}{0}{m+n}-2l N_1^s(m,n)\left(\Q{l}{s+1}{1}{m+n}+\frac{s}{2}\Q{l-1}{s}{0}{m+n}\right)\\
	&\quad -2N_2^s(m,n)\left((s-1-l)(s-1+l)\Msq\Q{l}{s-1}{0}{m+n}+l\S \Q{l-1}{s-1}{0}{m+n}\right)\,,
	\end{split}\\
	\begin{split}
	\acomm{\Q{1\vphantom{l}}{s}{1\vphantom{\xi}}{m}}{P_n}&=2\Q{1}{s+1}{1}{m+n}+\Q{0}{s}{0}{m+n}-2N_1^{s-1}(m,n)\left(\Msq\Q{1\vphantom{l}}{s-1}{0\vphantom{\xi}}{m+n}-\S\Q{0}{s-1}{0}{m+n}\right)\\
	&\quad -(s-1)\Msq N_2^{s-1}(m,n)\left(2(s-3)\Q{1}{s-1}{1}{m+n}-(s-2)\Q{0\vphantom{l}}{s-2}{0\vphantom{\xi}}{m+n}\right)\,,
	\end{split}
\end{align}
\end{subequations}
to derive a set of equations for $\c{l}{s}{\xi}{m}(u,r,\phi)$\,. The definition of the mode functions $N^s_i(m,n)$ is given in (\ref{def_capitalN}). The anti-commutation relation (\ref{middle_acomm}) is already restricted to $l\in\{0,1\}$\,. The resulting set of equations may be found in Appendix~\ref{appsub_spin2}.

This set of equations may be evaluated for specific index combinations and, e.g., for $c\equiv\c{0}{1}{0}{0}$ gives
\begin{subequations}
\begin{align}
	0&=\partial_{\!u}c-\frac{\Msq}{3}\left(4\c{0}{2}{0}{-1}-M(\phi)\c{0}{2}{0}{1}-4\c{1}{3}{1}{-1}+M(\phi)\c{1}{3}{1}{1}\right)-\frac{\S}{3}\left(4\c{1}{2}{0}{-1}-M(\phi)\c{1}{2}{0}{1}\right)\,,\\
	0&=\partial_{\!r}c-\frac{2}{3}\left(\Msq\c{0}{2}{0}{1}+\S\c{1}{2}{0}{1}-\Msq\c{1}{3}{1}{1}\right)\,,\\
	0&=\partial_{\!\phi}c+\frac{2\Msq}{3}\left(r \c{0}{2}{0}{0}+N(u,\phi)\c{0}{2}{0}{1}-r\c{1}{3}{1}{0}-N(u,\phi)\c{1}{3}{1}{1}\right)+\frac{2\S}{3}\left(r\c{1}{2}{0}{0}+N(u,\phi)\c{1}{2}{0}{1}\right)\,,
\end{align}
\end{subequations}
where we have omitted the coordinate dependence of the fields for the sake of better readability. The construction of a linear second-order partial differential equation for $c$, i.e. taking the ansatz
\begin{align}
    \left(\alpha+\beta^{\mu}\partial_{\!\mu}+\gamma^{\mu\nu}\partial_{\!\mu}\partial_{\nu}\right)c=0
\end{align}
and eliminating all derivatives by utilizing the respective partial differential equations admits only one non-trivial solution, which is
\begin{align}
    \left(\Box^{(2)}-4\Msq\right)c=0\,,
\end{align}
where the d'Alembert operator reads
\begin{align}
    \Box^{(2)}=\left(-M+\frac{N^2}{r^2}\right)\partial^2_{\!r}+\frac{\partial_{\!\phi}^2}{r^2}-2\partial_{\!u}\partial_{\!r}+\frac{2N}{r^2}\partial_{\!r}\partial_{\!\phi}+\left(-M-\frac{N^2}{r^2}+\frac{\partial_{\!\phi}N}{r}\right)\frac{\partial_{\!r}}{r}-\frac{\partial_{\!u}}{r}-\frac{N}{r^3}\partial_{\!\phi}\,.
\end{align}
This is precisely the d'Alembertian that arises from the metric (\ref{EF_metric})\,. So, the scalar field $c$, i.e. the unit component of the matter field $C$ (in this particular expansion), satisfies a Klein-Gordon equation with mass $2\mathcal{M}$\,.

We would like to conclude this sub-section with a few further remarks concerning the specific choice of $\ihs^{\text{\tiny (R)}}(\Msq,\S)$ as the underlying structure. The attempt to get a correspondence to the classical Klein-Gordon equation within the full algebra $\ihs(\Msq,\S)$ fails due to the appearance of coefficients $\c{s-1}{s}{s-1}{0}$\, that belong to the generators $\Q{s-1}{s}{s-1}{0}$, which is an index combination that only exists for $s$ being an odd number. Thus, in principle it would be sufficient to find a consistent algebra structure that does not contain these generators. We chose to restrict ourselves to what we call $\ihs^{\text{\tiny (R)}}(\Msq,\S)$ only because of its much simpler structure and it's ability to provide us with a consistent working example.
\subsection{Spin-3 Gauge Field and Generalized Klein-Gordon Equation}\label{spin3_eom}
We study the following spin-3 deformation, inspired by the gauge field considered in \cite{Afshar:2013vka} and in analogy to the chiral deformation (\ref{chiral_deform}) in the AdS case:
\begin{subequations}
\begin{align}
	\omega &=\left(J_1-\frac{M(\phi)}{4}J_{-1}-\frac{Z}{2} J_{-1}P_{-1}\right)d\!\phi\,,\\
    e&=\left(P_1-\frac{M(\phi)}{4}P_{-1}-\frac{Z}{4}P_{-1}P_{-1}\right)d\!u+\frac{1}{2}P_{-1}d\!r+\left(r P_0-\frac{N(u,\phi)}{2}P_{-1}\right)d\!\phi\,,
\end{align}
\end{subequations}
where $Z$ is a constant spin-3 field and $J_{-1}P_{-1}=\Q{1}{3}{0}{-2}$ as well as $P_{-1}P_{-1}=\Q{0}{3}{0}{-2}$\,. The zero-flatness and zero-torsion conditions are fulfilled by this gauge field. The necessary additional (anti-)commutation relations can again be obtained from (\ref{right_slice_multipl}) or (\ref{mult_Q_spin2}) and are presented in (\ref{spins_comm_acomm}) of the appendix, where they are included as special cases $\sigma=2$. 

The system of equations emerging from this set-up will not be displayed here but is included as the special case $\tilde{s}=3$ in Appendix~\ref{appsub_spins}. However, taking a generic ansatz for a linear third-order partial differential equation for $c$,
\begin{align}
    \left(\alpha+\beta^{\mu}\partial_{\!\mu}+\gamma^{\mu\nu}\partial_{\!\mu}\partial_{\nu}+\delta^{\mu\nu\rho}\partial_\mu\partial_\nu\partial_{\!\rho}\right)c=0\,,
\end{align}
and in addition setting $\S=0$ admits only one non-trivial solution:
\begin{align}
	\left(\Box^{(3)}-4\Msq\right)c=0\,,
\end{align}
where the spin-3 generalization of the d'Alembert operator is given by
\begin{align}
	\Box^{(3)}=\Box^{(2)}+\frac{3Z}{2r}\partial_r^2+Z \partial^3_{\!r}\,.
\end{align}
That is, in a spin-3 background the classical Klein-Gordon equation gets modified by additional contributions of $r$-derivatives up to third order which is similar to what happens in the AdS case.

The fact that we had to set the Casimir $\S$ to zero appears intriguing, though. At the current stage of our analysis it is not completely clear to us if there is a deeper physical reason behind this or not. One possible explanation for this might be our restriction to the right-slice sub-structure $\ihs^{\text{\tiny (R)}}(\Msq,\S)$ due to our incomplete knowledge of the complete algebra structure of $\ihs(\Msq,\S)$. It is also perceivable that by using a different sub-structure one does not necessarily have to set $\S$ to zero. This is something that we leave for future work.
\subsection{Klein-Gordon Equation in an Arbitrary-Spin Background}\label{spins_KG}
In order to generalize the construction we presented in the previous sub-section to a background that contains higher-spin charges with arbitrary spin $s$ we consider the gauge fields
\begin{subequations}
\begin{align}
    \omega &=\left(J_1-\sum_{\sigma=2}^{s}\frac{(\sigma-1)Z^{(\sigma)}}{4}J_{-1}(P_{-1})^{\sigma-2}\right)d\!\phi\,,\\
    e&=\left(P_1-\sum_{\sigma=2}^{s}\frac{Z^{(\sigma)}}{4}(P_{-1})^{\sigma-1}\right)d\!u+\frac{1}{2}P_{-1}d\!r+\left(r P_0-\frac{N(u,\phi)}{2}P_{-1}\right)d\!\phi\,,
\end{align}
\end{subequations}
where we have included one of the spin-2 fields as $Z^{(2)}\equiv M(\phi)$. All other fields $Z^{(\sigma)}$ for $\sigma\ge 3$ are taken to be constant. One may easily convince oneself that these fields obey the zero-flatness and zero-torsion conditions. In what follows we will again take the $\S$-Casimir to be zero, $\S\stackrel{!}{=}0$\,.

Expressions for the (anti-)commutation relations necessary to evaluate the equations of motion can be found in Appendix~\ref{app_strCons_right}, equations (\ref{spins_comm_acomm}). Expanding the matter field into generators exactly the same way as before and inserting those (anti-)commutation relations, we find a set of equations for the coefficients $\c{l}{s}{\xi}{m}$ within the master field $C$ that we display in Appendix~\ref{app_eom}. Therein we have relabelled the highest occurring spin to $\tilde{s}$ for the moment to avoid confusion with the general set of indices $(l,s,\xi,m)$\,. In particular, the equations for the scalar field $c\equiv\c{0}{1}{0}{0}$ read
\begin{subequations}
\begin{align}
    0&=\partial_{\!u}c-\frac{4\Msq}{3}\left(\c{0}{2}{0}{-1}-\c{1}{3}{1}{-1}\right)+\sum_{\sigma=2}^{\tilde{s}}\frac{(-1)^{\sigma}\mathcal{M}^{2(\sigma-1)}Z^{(\sigma)}}{4}\frac{(\sigma-1)!}{(\sigma-\sfrac{1}{2})^{\underline{\sigma-1}}}\left(2\c{0}{\sigma}{0}{\sigma-1}-\sigma \c{1}{\sigma+1}{1}{\sigma-1}\right)\,,\\
0&=\partial_{\!r}c-\frac{2\Msq}{3}\left(\c{0}{2}{0}{1}-\c{1}{3}{1}{1}\right)\,,\\
    0&=\partial_{\!\phi}c+\frac{2r\Msq}{3}\left(\c{0}{2}{0}{0}-\c{1}{3}{1}{0}\right)+\frac{2N(u,\phi)\Msq}{3}\left(\c{0}{2}{0}{1}-\c{1}{3}{1}{1}\right)\,.
\end{align}
\end{subequations}
While assembling the spin-2 d'Alembert operator acting on $c$ out of those equations, nearly all terms cancel out. The remaining sums can be identified after realizing that
\begin{align}
    \partial_{\!r}^k \c{0}{s}{0}{s-1}&=\frac{s^{\overline{k}}\mathcal{M}^{2k}}{(s+\sfrac{1}{2})^{\overline{k}}}\c{0}{s+k}{0}{s+k-1}-\frac{k(s+k)!(s-\sfrac{1}{2})^{\underline{s}}\mathcal{M}^{2k}}{2(s+k-1)(s-1)!(s+k-\sfrac{1}{2})^{\underline{s+k}}}\c{1}{s+k+1}{1}{s+k-1}\,.
\end{align}
The final equation, i.e. the generalization of the Klein-Gordon equation for a background deformed by an arbitrary number of higher-spin potentials $Z^{(\sigma)}=\const$ ($\sigma=3,\dots,s$), reads
\begin{align}
    \left(\Box^{(2)}-\sum_{\sigma=3}^s (-1)^{\sigma}Z^{(\sigma)}\left(\frac{\sigma}{2r}\partial_{\!r}^{\sigma-1}+\partial_{\!r}^{\sigma}\right)-4\Msq\right)=0 \,.
\end{align}
Similar to the spin-3 case that we treated earlier we see that also for general spin $s$ the d'Alembert operator gets modified by additional $r$-derivatives up to spin $s$. This result concludes our considerations.
\section{Summary and Outlook}
In this paper we discussed the formulation of three-dimensional flat-space higher-spin gravity based on a UEA construction. The guiding principle was to define a higher-spin theory of gravity as the Chern-Simons theory living on a quotient (generated by the Casimir elements) of the UEA of the classical isometry algebra. 

We summarize our main results:
\begin{enumerate}
    \item The UEA construction over the inhomogeneous Lie algebra $\isl(2,\R)$ yields an associative algebra that we propose to call $\ihs(\Msq,\S)$, which does not decompose into commuting parts as it is the case for its AdS companion $\hs(\lambda)\oplus\hs(\lambda)$. We derived the spin-$s$-spin-2 multiplication rules $\Q{l}{s}{\xi}{m}\star\Q{k}{2}{0}{n}$ and $\Q{k}{2}{0}{n}\star\Q{l}{s}{\xi}{m}$ as well as the structure constants of what we call the right slice $\ihs^{\text{\tiny (R)}}(\Msq,\S)$ consisting of generators $\Q{0}{s}{0}{m}$ and $\Q{1}{s}{\xi}{m}$.
    \item A straightforward contraction from $\hs(\lambda)\oplus\hs(\lambda)$ leads to an algebra that is related to a truncated version of the UEA construction but does not reproduce the full $\ihs(\Msq,\S)$. In particular, considering the direct sum of higher-spin algebras on the AdS side -- i.e. identifying mixed products with zero, $L_n\bar{L}_m\sim 0$, -- corresponds to a truncation of powers of translational generators on the flat-space side, i.e. $P_mP_n\sim 0$. The latter removes both the flat-space Casimir elements $\Msq$ and $\S$ from the theory.
    \item A connection between asymptotically AdS and asymptotically flat space in terms of an \.In\"on\"u-Wigner contraction that is compatible with a UEA construction on both the AdS and the flat-space side should be drawn from the larger algebra structure emerging from $\mathcal{U}(\sl(2,\R)\oplus\sl(2,\R))$. The prescription then involves higher powers of the contraction parameter $\epsilon$ (i.e. higher powers of the inverse length scale).
    \item We proposed the equation
    \begin{align}\label{eom_discussion}
        0&=\d C+\comm{\omega}{C}_{\!\star}+\acomm{e}{C}_{\!\star}
    \end{align}
    to appropriately describe the coupling of higher-spin fields and matter fields at linear order. The equation is inspired by linearized Vasiliev theory in AdS. We further assumed that the gauge fields $\omega$ and $e$ as well as the matter field $C$ are valued in (part of) the associative algebra $\ihs(\Msq,\S)$ and that we therefore can identify the star product with the algebra product of $\ihs(\Msq,\S)$.
    \item We sorted out the difference between the higher-spin Lie algebra governing the gauge sector and the associative structure necessary for matter-gravity coupling. We derived the structure constants of Lie-$\ihs$  and thus provided the definition of all finite-spin Lie algebras $\isl(N,\R)$.
    \item Expanding the matter field $C$ appearing in (\ref{eom_discussion}) within the sub-structure $\ihs^{\text{\tiny (R)}}(\Msq,\S)$ of $\ihs(\Msq,\S)$ is sufficient to describe the linearized coupling of a scalar field to a gauge background that is charged under an infinite number of higher-spin fields. We considered higher-spin gauge fields
    \begin{subequations}
    \begin{align}
        \omega &=\left(J_1-\sum_{\sigma=2}^{s}\frac{(\sigma-1)Z^{(\sigma)}}{4}J_{-1}(P_{-1})^{\sigma-2}\right)d\!\phi\,,\\
        e&=\left(P_1-\sum_{\sigma=2}^{s}\frac{Z^{(\sigma)}}{4}(P_{-1})^{\sigma-1}\right)d\!u+\frac{1}{2}P_{-1}d\!r+\left(r P_0-\frac{N(u,\phi)}{2}P_{-1}\right)d\!\phi\,,
    \end{align}
    \end{subequations}
    where $Z^{(\sigma)}$ are constant higher-spin charges, except for $\sigma=2$, where $Z^{(2)}\equiv M(\phi)$, and found the scalar field $c$ of mass $m$ (appearing as the coefficient of the unit element in the expansion of $C$) to fulfil the generalized Klein-Gordon equation
    \begin{align}
    \left(\Box^{(2)}-\sum_{\sigma=3}^s (-1)^{\sigma}Z^{(\sigma)}\left(\frac{\sigma}{2r}\partial_{\!r}^{\sigma-1}+\partial_{\!r}^{\sigma}\right)\right)c=m^2 c
    \end{align}
    in this background, where the mass is connected to the parametrization of the $\isl(2,\R)$-Casimir element by $m=2\mathcal{M}$\,. Including deformations of unbounded spin is simply achieved by letting $s\rightarrow\infty$. The classical spin-2 case is included for vanishing higher-spin charges.
\end{enumerate}

Even though we were able to work out the general structure of $\ihs(\Msq,\S)$ there is quite an amount of open tasks concerning the detailed structure of this algebraic object. First of all, it would be nice to have full control over the multiplication rules of arbitrary generators, i.e. to find a closed-form expression for the product $\Q{l}{s}{\xi}{m}\star\Q{k}{t}{\eta}{n}$\,. Up until now a full combinatorial treatment of these products is still work in progress.

In connection to that rather technical issue, one might want to allow for pairs of non-constant higher-spin charges, $Z^{(\sigma)}(\phi)$ and $W^{(\sigma)}(u,\phi)$, connected by integrability conditions, much like it was proposed in \cite{Afshar:2013vka} for the spin-3 case. Extending our analysis to also include higher-spin potentials and compare to the existing literature, such as \cite{Gary:2014ppa,Matulich:2014hea}, might be of interest. Having to restrict to only one set of constant charges might be due to the fact that we expanded the master field within the right slice only. Intuitively we would like to use the full algebra $\ihs(\Msq,\S)$, except for generators $\Q{s-1}{s}{s-1}{0}$ (which, as we found, spoil the emergence of a classical KG equation in the first place). 

Our results can be taken as a starting point to explore several directions in three-dimensional flat-space higher-spin gravity. A pressing question concerns the role of the various no-go theorems and whether or not they can actually be circumvented by consideration of a theory with dynamically broken flat-space symmetry. In particular, a decent answer to that question would require us to find a framework that allows for a back-reaction of the matter fields to the higher-spin background.

Note that, though we were widely concerned with a higher-spin set-up, our results also contribute to the classical (spin 2) sector: the linear coupling of scalar fields to gravity by usage of an associative algebra product in the Chern-Simons formalism has not yet been discussed for the case of asymptotically flat space-times so far.

Some interesting technical directions consist in the construction of an asymptotic symmetry algebra connected to $\ihs(\Msq,\S)$-valued gauge fields as well as a generalization to a supersymmetric sector.

Finally, a propagating scalar field may be employed as a probe in the examination of correlation functions in flat-space holography when its dual operator is inserted on the asymptotic boundary.
\section*{Acknowledgements}
We would like to thank Nicolas Boulanger, Andrea Campoleoni, Daniel Grumiller, Stefan Prohazka and Mikhail Vasiliev for delightful scientific discussions and comments.

MA is funded by the Deutsche Forschungsgemeinschaft (DFG, German Research  Foundation) under Grant No.\,406235073 within the Heisenberg program.

The work of MP is funded by the Deutsche Forschungsgemeinschaft (DFG) under Grant No.\,406116891 within the Research Training Group RTG\,2522/1 as well as a Landesgraduiertenstipendium of the federal state of Thuringia. 

The research of MR is supported by the European Union’s Horizon 2020 research and innovation programme under the Marie Skłodowska-Curie grant agreement No. 832542 as well as the DOE grant de-sc/0007870.
%
%
%
%%%%%%%%%%%%%%%%%%%%%%%%%%%%%%%%%%%%%%%%%%%%%%%%%%%%%%%%%%%%%%%%%%%%%%%%%%
%%%%%%%%%%%%%%%%%%%%%%%%%%%%%%%%%%%%%%%%%%%%%%%%%%%%%%%%%%%%%%%%%%%%%%%%%%
%\newpage
\appendix
\section{Product Rules of the Algebra \texorpdfstring{$\ihs(\Msq,\S)$}{ihs(M\texttwosuperior,S)}}\label{app_largeUEA}
This appendix summarizes various multiplication rules and commutation relations we obtained in the study of $\ihs(\Msq,\S)$ and, in particular, its right slice. Furthermore, we would like to draw the reader's attention to a \emph{Mathematica} notebook, which we attached to the \emph{arXiv} version of this article, that provides an implementation of the formal product and contains some explicit checks of our results.
\subsection{Formal-Product Relations of \texorpdfstring{$\mathcal{U}(\isl(2,\R))$}{U(isl(2,R))}}\label{app_uea}
The product rules of this algebra can be fully derived by combinatorial considerations, see \cite{diaz2016combinatorics} for the respective construction in the case of $\sl(2,\R)$, which we are closely following in the present case. All relations needed to express disordered formal products in terms of ordered ones are:
{\small
\begin{subequations}\label{uea_StrC}
\begin{align}
    (J_{-1})^a(J_0)^b&=\left(J_0-a\right)^b(J_{-1})^a=\mathlarger{\sum}_{k=0}^b (-1)^k a^k \binom{b}{k}(J_0)^{b-k} (J_{-1})^a\,,\\
    (J_0)^a(J_1)^b&=(J_1)^b \mathlarger{\sum}_{k=0}^a(-1)^k b^k\binom{a}{k}(J_0)^{a-k}\,,\\
    (J_{-1})^a(J_1)^b&=\mathlarger{\sum}_{k=0}^{\min(a,b)}\mathlarger{\sum}_{j=0}^k (-2)^{k-j}k!\binom{a}{k}\binom{b}{k}\left(a+b-2k\right)^j_k (J_1)^{b-k}(J_0)^{k-j}(J_{-1})^{a-k}\,,\\
     (P_{\pm 1})^a(J_0)^b&=\left(J_0\pm a\right)^b(P_{\pm 1})^a\,,\phantom{\mathlarger{\sum}_{k=0}^a \binom{b}{k}}\\
    (P_0)^a(J_{-1})^b&=\mathlarger{\sum}_{k=0}^a k! \binom{a}{k}\binom{b}{k}(J_{-1})^{b-k}(P_0)^{a-k}(P_{-1})^k\,,\\
    (P_0)^a(J_1)^b&=\mathlarger{\sum}_{k=0}^b (-1)^k k! \binom{a}{k}\binom{b}{k} (J_1)^{b-k}(P_1)^k(P_0)^{a-k}\,,\\
    (P_{-1})^a(J_1)^b&=\mathlarger{\sum}_{k=0}^{\min(a,b)}\mathlarger{\sum}_{j=0}^k (-1)^{k+j}2^{k-j}k!j!\binom{a}{k}\binom{b}{k}\binom{k}{j}\binom{b-k}{j}(J_1)^{b-k-j}(P_1)^j(P_0)^{k-j}(P_{-1})^{a-k}			\,,\\
    (P_1)^a(J_{-1})^b&=\mathlarger{\sum}_{k=0}^{\min(a,b)}\mathlarger{\sum}_{j=0}^k 2^{k-j}k!j!\binom{a}{k}\binom{b}{k}\binom{k}{j}\binom{b-k}{j}(J_{-1})^{b-k-j}(P_1)^{a-k}(P_0)^{k-j}(P_{-1})^j\,,
\end{align}
\end{subequations}
}%
where we use the definition
\begin{align}
    (a)^s_n:=e^s_n(a,a+1,\dots,a+n-1)
\end{align}
in terms of the elementary symmetric function
\begin{align}
    e^s_n(x_1,\dots,x_n):=\mathlarger{\sum}_{1\le i_1 < .. < i_s \le n} x_{i_1} \dots x_{i_s}\,.
\end{align}
Naturally, the reverse ordering obeys quite similar rules as well, e.g.
{\small
\begin{subequations}
\begin{align}
    (J_{-1})^a(P_1)^b &=\mathlarger{\sum}_{k=0}^{\min(a,b)}\mathlarger{\sum}_{j=0}^k (-1)^{k+j} 2^{k-j}k!j!\binom{a}{k}\binom{b}{k}\binom{k}{j}\binom{a-k}{j}(P_1)^{b-k}(P_0)^{k-j}(P_{-1})^j(J_{-1})^{a-k-j}\,,\\
   (J_1)^a(P_{-1})^b &=\mathlarger{\sum}_{k=0}^{\min(a,b)}\mathlarger{\sum}_{j=0}^k 2^{k-j}k!j!\binom{a}{k}\binom{b}{k}\binom{k}{j}\binom{a-k}{j}(P_1)^{j}(P_0)^{k-j}(P_{-1})^{b-k}(J_{1})^{a-k-j}\,.
\end{align}
\end{subequations}
}%
\subsection{Products of Spin-\texorpdfstring{$s$}{s} and Spin-2 Generators}
We give a short summary of multiplication rules that we obtained in the full higher-spin algebra $\ihs(\Msq,\S)$. The generators of the algebra are denoted $\Q{l}{s}{\xi}{m}$ and the range of indices is illustrated in table \ref{table_of_generators} up to $s=5$. Though we use ``$\star$'' as notation for the product of two such generators it can always be translated to the formal product of powers of $\isl(2,\R)$-generators. The star-notation only indicates a specific choice of basis in that sense.
\begin{table}[h!]
\caption{The allowed range of indices of the generators $\Q{l}{s}{\xi}{m}$ of the algebra $\ihs(\Msq,\S)$ depicted up to spin $s=5$.}
\label{table_of_generators}
\begin{center}
\scalebox{0.665}{
\begin{tabular}{cc@{\hskip 35pt}ccc@{\hskip 35pt}cccc@{\hskip 35pt}ccccccc}
\toprule
\multicolumn{2}{c}{\textbf{Spin 2}}\hspace{40pt}  & \multicolumn{3}{c}{\textbf{Spin 3}}\hspace{40pt}	& \multicolumn{4}{c}{\textbf{Spin 4}}\hspace{40pt}		& \multicolumn{5}{c}{\textbf{Spin 5}}\hspace{30pt}	& 	& \\
$l=1$		& $l=0$					  & $l=2$	& $l=1$		& $l=0$			& $l=3$		& $l=2$	& $l=1$		& $l=0$		& $l=4$		& $l=3$		& $l=2$	& $l=1$		& $l=0$ & \\
\cmidrule{1-14}
$\Q{1}{2}{0}{m}$ & $\Q{0}{2}{0}{m}$ & $\Q{2}{3}{0}{m}$ & $\Q{1}{3}{0}{m}$ & $\Q{0}{3}{0}{m}$ & $\Q{3}{4}{0}{m}$ & $\Q{2}{4}{0}{m}$ & $\Q{1}{4}{0}{m}$ & $\Q{0}{4}{0}{m}$ & $\Q{4}{5}{0}{m}$ & $\Q{3}{5}{0}{m}$ & $\Q{2}{5}{0}{m}$ & $\Q{1}{5}{0}{m}$ & $\Q{0}{5}{0}{m}$ & &\ $|m|\le s-1$\\
\addlinespace
 & & & $\Q{1}{3}{1}{m}$ & & & $\Q{2}{4}{1}{m}$ & $\Q{1}{4}{1}{m}$ & & & $\Q{3}{5}{1}{m}$ & $\Q{2}{5}{1}{m}$ & $\Q{1}{5}{1}{m}$ & & &\ $|m|\le s-2$\\
 \addlinespace
 & & $\Q{2}{3}{2}{m}$ & & & $\Q{3}{4}{2}{m}$ & $\Q{2}{4}{2}{m}$ & & & $\Q{4}{5}{2}{m}$ & $\Q{3}{5}{2}{m}$ & $\Q{2}{5}{2}{m}$ & & & &\ $|m|\le s-3$\\
 \addlinespace
 & & & & & & & & & & $\Q{3}{5}{3}{m}$ & & & & &\ $|m|\le s-4$\\
 \addlinespace
 & & & & & & & & & $\Q{4}{5}{4}{m}$ & & & & & &\ $|m|\le s-5$\\
\bottomrule
\end{tabular}
}
\end{center}
\end{table}

The useful identity for the adjoint action of an algebra element $A$, applied $n$ times to another algebra object $B$,
\begin{align}
	\ad_A^n (B)=\sum_{k=0}^{n}(-1)^{n-k}\binom{n}{k}A^k B A^{n-k}\,,
\end{align}
allows us to expand the nested commutators in the definition of our generators into formal powers of $\isl(2,\R)$-generators,
\begin{align}\label{descendant_expansion}
	\Q{l}{s}{\xi}{m}&=\frac{(s+m-\xi-1)!}{(2s-2\xi-2)!}\sum_{k=0}^{s-1-\xi-m} (-1)^k \binom{s-1-\xi-m}{k} (J_{-1})^k\ \Q{l}{s}{\xi}{s-1-\xi}\ (J_{-1})^{s-1-\xi-m-k}\,,
\end{align}
which may be used as a starting point for the calculation of product rules by exploiting our knowledge of the algebra $\mathcal{U}(\isl(2,\R))$ presented in sub-section \ref{app_uea}\,. Here, only the results will be given.

We use the compact notation $a^{\overline{k}}$ and $a^{\underline{k}}$ for the rising and falling factorial, respectively. Furthermore, we define the abbreviations
\begin{subequations}\label{def_capitalN}
\begin{align}
	N^s_1(m,n)&:=\frac{m-(s-1)n}{s^{\underline{2}}}\,,\\
	N^s_2(m,n)&:=\frac{m^2+(s-1)(2s-3)n^2-(2s-3)mn-(s-1)^2}{(s-1)^2(2s-1)(2s-3)}\,,
\end{align}
\end{subequations}
as well as the coefficients
\begin{subequations}
\begin{align}
	\begin{split}
	\alpha^n_k &= \frac{2^k k}{(2k)!}\frac{(n+k-2)!}{(n-k)!}\left((k-1)\,\F{-(k-1)}{-(n-k)}{-(n+k-2)}{-1}\right.\\
	&\qquad\qquad\qquad\qquad\quad \left.+(n+k-1)\F{-k}{-(n+1-k)}{-(n+k-1)}{-1}\right)\,,
	\end{split}\\[0.2cm]
	\beta^n_k &= 2^k\binom{n+k}{2k}\F{-k}{-(n-k)}{-(n+k)}{-1}\,,\\[0.2cm]
	\gamma^n_k &= \begin{cases}
        2^k\frac{(3k-3)^{\overline{n-k}}}{(n-k)!}\tensor*[_{3\!}]{F}{_2}\left[\left.\begin{matrix}-(k-2)\,, & -\frac{n-1-k}{2}\,, & -\frac{n-k}{2}\\ -\frac{n}{2}-k+2\,, &\multicolumn{2}{c}{-\frac{n-1}{2}-k+2}  \end{matrix}\ \right| 1\right]\,, & k>1\\
        0\,, & k=1
    \end{cases}\ \
\end{align}
\end{subequations}
for $n\ge 0$ and $0\le k\le n$. Note that $\alpha^n_0=0=\gamma^n_0$ and $\beta^n_0=1$. The origin of these coefficients lies in the commutation rules of generators $P_n$ with powers of the element $\C$. In particular,
\begin{subequations}
\begin{align}
    P_{\pm 1}\C^n&=\sum_{k=0}^{n} \C^{n-k}\left[\pm\alpha^n_k\left(J_0P_{\pm 1}-J_{\pm 1}P_0\right)+\beta^n_k P_{\pm 1}-\gamma^n_k \S J_{\pm 1}\right]\,,\\
    P_0\C^n&=\sum_{k=0}^{n} \C^{n-k}\left[\alpha^n_k\left(J_0P_0-J_{1}P_{-1}-\S\right)+\beta^n_k P_0-\gamma^n_k \S J_0\right]\,.
\end{align}
\end{subequations}

We can now write down the following product relations between general basis elements of our algebra and spin-2 generators $\Q{1}{2}{0}{n}=J_n$\,, $\Q{0}{2}{0}{n}=P_n$. We necessarily need to distinguish the cases of $\xi$ being an even or an odd number:
% \begingroup
% \allowdisplaybreaks
{%
\begingroup
\allowdisplaybreaks\scriptsize
\begin{subequations}\label{mult_Q_spin2}
\begin{align}
\begin{split}
	\Q{l}{s}{\xi}{m}\star J_{n}\at{4}{1.5}{\xi\text{ even}}&=\Q{l+1}{s+1}{\xi}{m+n}+N_1^{s-\xi}(m,n)\left[(s-1-l)\Q{l+1}{s+1}{\xi+1}{m+n}-\frac{(l-\xi)(2s-1-l-\xi)-2(s-\xi)^{\underline{2}}}{2}\Q{l}{s}				{\xi}{m+n}\right]\\
    &\quad -N_2^{s-\xi}(m,n)\left[(l-\xi)(2s-2-l-\xi)\Q{l+1}{s+1}{\xi+2}{m+n}+(l-\xi)(2s-2-l-\xi)(s-1-l)\Q{l}{s}{\xi+1}{m+n}\right.\\
    &\quad \left.-(s-1-l)^{\underline{2}}\Msq\ \Q{l+1}{s-1}{\xi}{m+n}+(s-1-l)(2s-2l-3)\S\Q{l}{s-1}{\xi}{m+n}-\frac{(l-\xi)^{\underline{2}}(2s-1-l-\xi)^{\underline{2}}}{4}\Q{l-1}{s-1}{\xi}{m+n}\right]\,,
\end{split}\\
\begin{split}
     \Q{l}{s}{\xi}{m}\star J_{n}\at{4}{1.5}{\xi\text{odd}}&=\Q{l+1}{s+1}{\xi}{m+n}+N_1^{s-\xi}(m,n)\left[(s-1-l)\Q{l+1}{s+1}{\xi+1}{m+n}-\frac{(l+1-\xi)(2s-2-l-\xi)-2(s-\xi)^{\underline{2}}}{2}\Q{l}{s}			{\xi}{m+n}\right.\\
    &\quad \left.+(s-2-l)\Msq\ \Q{l+1}{s-1}{\xi-1}{m+n}-(2s-2l-3)\S\ \Q{l}{s-1}{\xi-1}{m+n}\right]-N_2^{s-\xi}(m,n)\left[(l-\xi)(2s-2-l-\xi)\Q{l+1}{s+1}{\xi+2}{m+n}\right.\\
    &\quad \left.+(l-\xi)(2s-2-l-\xi)(s-1-l)\Q{l}{s}{\xi+1}{m+n}-(s-2-l)^{\underline{2}}\Msq\ \Q{l+1}{s-1}{\xi}{m+n}\right.\\
    &\quad \left.+(s-2-l)(2s-2l-3)\S\Q{l}{s-1}{\xi}{m+n}-\frac{(l+1-\xi)^{\underline{2}}(2s-2-l-\xi)^{\underline{2}}}{4}\Q{l-1}{s-1}{\xi}{m+n} \right.\\
    &\quad +(l+1-\xi)(s-2-l)(2s-l-\xi-3)\Msq\ \Q{l}{s-2}{\xi-1}{m+n}\\
    &\quad \left. +(2s-2l-3)\left(s(2\xi-1)-\xi(\xi+2)-l(2s-3)+l^2+1\right)\S\ \Q{l-1}{s-2}{\xi-1}{m+n}\right]\,,
\end{split}\\
\begin{split}
	J_{n}\star\Q{l}{s}{\xi}{m}\at{4}{1.5}{\xi\text{ even}}&=\Q{l+1}{s+1}{\xi}{m+n}+N_1^{s-\xi}(m,n)\left[(s-1-l)\Q{l+1}{s+1}{\xi+1}{m+n}-\frac{(l-\xi)(2s-1-l-\xi)}{2}\Q{l}{s}{\xi}{m+n}\right]\\
    &\quad -N_2^{s-\xi}(m,n)\left[(l-\xi)(2s-2-l-\xi)\Q{l+1}{s+1}{\xi+2}{m+n}+(l-\xi)(2s-2-l-\xi)(s-1-l)\Q{l}{s}{\xi+1}{m+n}\right.\\
    &\quad \left.-(s-1-l)^{\underline{2}}\Msq\ \Q{l+1}{s-1}{\xi}{m+n}+(s-1-l)(2s-2l-3)\S\Q{l}{s-1}{\xi}{m+n}-\frac{(l-\xi)^{\underline{2}}(2s-1-l-\xi)^{\underline{2}}}{4}\Q{l-1}{s-1}{\xi}{m+n}\right]\,,
\end{split}\\
\begin{split}
    J_{n}\star\Q{l}{s}{\xi}{m}\at{4}{1.5}{\xi\text{ odd}}&=\Q{l+1}{s+1}{\xi}{m+n}+N_1^{s-\xi}(m,n)\left[(s-1-l)\Q{l+1}{s+1}{\xi+1}{m+n}-\frac{(l+1-\xi)(2s-2-l-\xi)}{2}\Q{l}{s}{\xi}{m+n}\right.\\
    &\quad \left.+(s-2-l)\Msq\ \Q{l+1}{s-1}{\xi-1}{m+n}-(2s-2l-3)\S\ \Q{l}{s-1}{\xi-1}{m+n}\right]\\
    &\quad -N_2^{s-\xi}(m,n)\left[(l-\xi)(2s-2-l-\xi)\Q{l+1}{s+1}{\xi+2}{m+n}+(l-\xi)(2s-2-l-\xi)(s-1-l)\Q{l}{s}{\xi+1}{m+n}\right.\\
    &\quad -(s-2-l)^{\underline{2}}\Msq\ \Q{l+1}{s-1}{\xi}{m+n}+(s-2-l)(2s-2l-3)\S\Q{l}{s-1}{\xi}{m+n}\\
    &\quad \left.-\frac{(l+1-\xi)^{\underline{2}}(2s-2-l-\xi)^{\underline{2}}}{4}\Q{l-1}{s-1}{\xi}{m+n} +(l+1-\xi)(s-2-l)(2s-l-\xi-3)\Msq\ \Q{l}{s-2}{\xi-1}{m+n}\right.\\
    &\quad \left.+(2s-2l-3)\left(s(2\xi-1)-\xi(\xi+2)-l(2s-3)+l^2+1\right)\S\ \Q{l-1}{s-2}{\xi-1}{m+n}\right]\,;
\end{split}\\
\begin{split}
	\Q{l}{s}{\xi}{m}\star P_{n}\at{4}{1.5}{\xi\text{ even}}&=\Q{l}{s+1}{\xi}{m+n}-N^{s-\xi}_1(m,n)(l-\xi)\left(\Q{l}{s+1}{\xi+1}{m+n}-\frac{l-\xi-1}{2}\Q{l-1}{s}{\xi}{m+n}\right)+N^{s-\xi}_2(m,n)\left((l-\xi)^{\underline{2}}\Q{l}{s+1}{\xi+2}{m+n}\right.\\
    &\quad \left.-\frac{(l-\xi)^{\underline{2}}(2l-2\xi-1)}{2}\Q{l-1}{s}{\xi+1}{m+n}-(s-1-l)(s-1+l-2\xi)\Msq\Q{l}{s-1}{\xi}{m+n}\right.\\
    &\quad\left.-(l-\xi)(2l-2\xi-1)\S\Q{l-1}{s-1}{\xi}{m+n}+\frac{(l-\xi)^{\underline{3}}(l-\xi-1)}{4}\Q{l-2}{s-1}{\xi}{m+n}\right)\,,
\end{split}\\
\begin{split}
\Q{l}{s}{\xi}{m}\star P_{n}\at{4}{1.5}{\xi\text{ odd}}&=\Q{l}{s+1}{\xi}{m+n}-N^{s-\xi}_1(m,n)\left((l-\xi)\Q{l}{s+1}{\xi+1}{m+n}-\frac{(l+1-\xi)^{\underline{2}}}{2}\Q{l-1}{s}{\xi}{m+n}+\Msq(l+1-\xi)\Q{l}{s-1}{\xi-1}{m+n}\right.\\
    &\quad\left. -\S(2l-2\xi+1)\Q{l-1}{s-1}{\xi-1}{m+n}\right)+N^{s-\xi}_2(m,n)\left((l-\xi)^{\underline{2}}\Q{l}{s+1}{\xi+2}{m+n}-\frac{(l-\xi)^{\underline{2}}(2l-2\xi-1)}{2}\Q{l-1}{s}{\xi+1}{m+n}\right.\\
    &\quad\left.-(s-2-l)(s+l-2\xi)\Msq\Q{l}{s-1}{\xi}{m+n}-(l-\xi)(2l-2\xi+1)\S\Q{l-1}{s-1}{\xi}{m+n}+\frac{(l+1-\xi)^{\underline{3}}(l-\xi)}{4}\Q{l-2}{s-1}{\xi}{m+n}\right.\\
    &\quad\left.-\Msq\frac{(l+1-\xi)^{\underline{2}}(2l-2\xi+1)}{2}\Q{l-1}{s-2}{\xi-1}{m+n}+\S(l-\xi)(2(l-\xi)^2-1)\Q{l-2}{s-2}{\xi-1}{m+n}\right)\,,
\end{split}\\
\begin{split}
	P_n\star\Q{l}{s}{\xi}{m}\at{4}{1.5}{\xi\text{ even}}&=\sum_{j=0}^{\frac{\xi}{2}}\left\{\alpha^{\frac{\xi}{2}}_{\!j} \Q{l+1-2j}{s+2-2j}{\xi+1-2j}{m+n}+\beta^{\frac{\xi}{2}}_{\!j} \Q{l-2j}{s+1-2j}{\xi-2j}{m+n}-\S \gamma^{\frac{\xi}{2}}_{\!j} \Q{l+1-2j}{s+1-2j}{\xi-2j}{m+n}\right.\\
    &\quad -N^{s-\xi}_1(m,n)\left[(l-\xi)\alpha^{\frac{\xi}{2}}_{\!j} \Q{l+1-2j}{s+2-2j}{\xi+2-2j}{m+n} +(s-1-l)\S\gamma^{\frac{\xi}{2}}_{\!j} \Q{l+1-2j}{s+1-2j}{\xi+1-2j}{m+n}\right.\\
    &\quad +\frac{l-\xi}{2}\left((2s-1-l-\xi)\alpha^{\frac{\xi}{2}}_{\!j}+2\beta^{\frac{\xi}{2}}_{\!j}\right) \Q{l-2j}{s+1-2j}{\xi+1-2j}{m+n}-(s-1-l)\Msq\alpha^{\frac{\xi}{2}}_{\!j} \Q{l+1-2j}{s-2j}{\xi-2j}{m+n}\\
    &\quad +\S\left((s-1-2l+\xi)\alpha^{\frac{\xi}{2}}_{\!j}-\frac{(l-\xi)(2s-1-l-\xi)}{2}\gamma^{\frac{\xi}{2}}_{\!j}\right) \Q{l-2j}{s-2j}{\xi-2j}{m+n}\\
    &\quad \left.+\frac{(l-\xi)(2s+1-l-\xi)}{2}					\beta^{\frac{\xi}{2}}_{\!j} \Q{l-1-2j}{s-2j}{\xi-2j}{m+n}\right]\\
    &\quad +N^{s-\xi}_2(m,n)\left[(l-\xi)^{\underline{2}}\ \alpha^{\frac{\xi}{2}}_{\!j} \Q{l+1-2j}{s+2-2j}{\xi+3-2j}{m+n}+(l-\xi)(2s-2-l-\xi)\S\gamma^{\frac{\xi}{2}}_{\!j} \Q{l+1-2j}{s+1-2j}{\xi+2-2j}			{m+n}\right.\\
    &\quad +\frac{(l-\xi)^{\underline{2}}}{2}\left((4s-2l-2\xi-1)\alpha^{\frac{\xi}{2}}_{\!j}+2\beta^{\frac{\xi}{2}}_{\!j}\right) \Q{l-2j}{s+1-2j}{\xi+2-2j}{m+n}+(s-1-l)^{\underline{2}}\Msq					\alpha^{\frac{\xi}{2}}_{\!j} \Q{l+1-2j}{s-2j}{\xi+1-2j}{m+n}\\
    &\quad +(s-1-l)(l-\xi)\S\left(2\alpha^{\frac{\xi}{2}}_{\!j}+(2s-2-l-\xi)\gamma^{\frac{\xi}{2}}_{\!j}\right) \Q{l-2j}{s-2j}{\xi+1-2j}{m+n}+\frac{(l-\xi)^{\underline{2}}}{4}\left((2s-1-l-\xi)^{\underline{2}}\alpha^{\frac{\xi}{2}}_{\!j}\right.\\
    &\quad\left. +2(4s-2l-2\xi-1)\beta^{\frac{\xi}{2}}_{\!j}\right) \Q{l-1-2j}{s-2j}{\xi+1-2j}{m+n}-(s-1-l)^{\underline{2}}\Msq\S \gamma^{\frac{\xi}{2}}_{\!j} \Q{l+1-2j}{s-1-2j}{\xi-2j}{m+n}\\
    &\quad -(s-1-l)\left(\Msq\left((l-\xi)(2s-2-l-\xi)\alpha^{\frac{\xi}{2}}_{\!j}+(s-1+l-2\xi)\beta^{\frac{\xi}{2}}_{\!j}\right)-(2s-2l-3)\S^2\gamma^{\frac{\xi}{2}}_{\!j}\right) \Q{l-2j}{s-1-2j}{\xi-2j}{m+n}\\
    &\quad +\frac{l-\xi}{4}\S\left(2(2(s+\xi)(2s-\xi)+4l^2-5\xi-l(10s-2\xi-5)-3)\alpha^{\frac{\xi}{2}}_{\!j}-4(2l-2\xi-1)\beta^{\frac{\xi}{2}}_{\!j}\right.\\
    &\quad\left.\left.\left. -(l-1-\xi)(2s-1-l-\xi)^{\underline{2}}\ \gamma^{\frac{\xi}{2}}_{\!j}\right) \Q{l-1-2j}{s-1-2j}{\xi-2j}{m+n}+\frac{(l-\xi)^{\underline{2}}(2s+1-l-\xi)^{\underline{2}}}{4}\beta^{\frac{\xi}{2}}_{\!j} \Q{l-2-2j}{s-1-2j}{\xi-2j}{m+n}\right]\right\}\,,
\end{split}\\
\begin{split}
	P_n\star\Q{l}{s}{\xi}{m}\at{4}{1.5}{\xi\text{ odd}}&=\sum_{j=0}^{\left\lfloor\!\frac{\xi}{2}\!\right\rfloor}\left\{\StrCons{\alpha} \Q{l+1-2j}{s+2-2j}{\xi+1-2j}{m+n}+\StrCons{\beta} \Q{l-2j}{s+1-2j}{\xi-2j}{m+n}-\S \StrCons{\gamma} \Q{l+1-2j}{s+1-2j}{\xi-2j}{m+n}+\Msq\StrCons{\alpha} \Q{l+1-2j}{s-2j}{\xi-1-2j}{m+n}\right.\\
	&\quad -2\S\StrCons{\alpha} \Q{l-2j}{s-2j}{\xi-1-2j}{m+n}+\StrCons{\beta} \Q{l-1-2j}{s-2j}{\xi-1-2j}{m+n}-N^{s-\xi}_1(m,n)\left[(l-\xi)\StrCons{\alpha} \Q{l+1-2j}{s+2-2j}{\xi+2-2j}{m+n}\right.\\
    &\quad +(s-1-l)\S\StrCons{\gamma} \Q{l+1-2j}{s+1-2j}{\xi+1-2j}{m+n}+\frac{l-\xi}{2}\left((2s+1-l-\xi)\StrCons{\alpha}+2\StrCons{\beta}\right)\Q{l-2j}{s+1-2j}{\xi+1-2j}{m+n}\\
    &\quad -(s-2-l)\Msq\StrCons{\alpha}\Q{l+1-2j}{s-2j}{\xi-2j}{m+n}\\
    &\quad +\S\left((s-2-2l+\xi)\StrCons{\alpha}-\frac{(l+1-\xi)(2s-2-l-\xi)}{2}\StrCons\gamma\right)\Q{l-2j}{s-2j}{\xi-2j}{m+n}\\
    &\quad +\frac{(l-\xi)(2s+1-l-\xi)}{2}\StrCons{\beta}\Q{l-1-2j}{s-2j}{\xi-2j}{m+n}+(s-2-l)\Msq\S\StrCons{\gamma}\Q{l+1-2j}{s-1-2j}{\xi-1-2j}{m+n}\\
    &\quad +\left(\frac{(l+1-\xi)(2s-2-l-\xi)}{2}\Msq\StrCons{\alpha}+(l+1-\xi)\Msq\StrCons{\beta}-(2s-2l-3)\S^2\StrCons{\gamma}\right)\Q{l-2j}{s-1-2j}{\xi-1-2j}{m+n}\\
    &\quad -\S\left((l-\xi)(2s-1-l-\xi)\StrCons{\alpha}+(2l-2\xi+1)\StrCons{\beta}\right)\Q{l-1-2j}{s-1-2j}{\xi-1-2j}{m+n}\\
    &\quad \left.+\frac{(l-\xi)(2s+1-l-\xi)}{2}\StrCons{\beta}\Q{l-2-2j}{s-1-2j}{\xi-1-2j}{m+n}\right]+N^{s-\xi}_2(m,n)\left[(l-\xi)^{\underline{2}}\StrCons{\alpha}\Q{l+1-2j}{s+2-2j}{\xi+3-2j}{m+n}\right.\\
    &\quad +(l-\xi)(2s-2-l-\xi)\S\StrCons{\gamma}\Q{l+1-2j}{s+1-2j}{\xi+2-2j}{m+n}+(l-\xi)^{\underline{2}}\left((2s-l-\xi-\sfrac{1}{2})\StrCons{\alpha}+\StrCons{\beta}\right)\Q{l-2j}{s+1-2j}{\xi+2-2j}{m+n}\\
    &\quad +\Msq\left(s(s-3)-2l(s+\xi-2)+\xi^{\underline{2}}+2l^2+3\right)\StrCons{\alpha}\Q{l+1-2j}{s-2j}{\xi+1-2j}{m+n}\\
    &\quad +(l-\xi)\S\left(2(s+\xi-2l-1)\StrCons{\alpha}+(s-1-l)(2s-2-l-\xi)\StrCons{\gamma}\right)\Q{l-2j}{s-2j}{\xi+1-2j}{m+n}\\
    &\quad +\frac{(l-\xi)^{\underline{2}}}{4}\left((2s+1-l-\xi)^{\underline{2}}\StrCons{\alpha}+2(4s+1-2l-2\xi)\StrCons{\beta}\right)\Q{l-1-2j}{s-2j}{\xi+1-2j}{m+n} \\
    &\quad -(s-2-l)^{\underline{2}}\Msq\S\StrCons{\gamma}\Q{l+1-2j}{s-1-2j}{\xi-2j}{m+n}-(s-2-l)\left((l+1-\xi)(2s-3-l-\xi)\Msq\StrCons{\alpha}\right.\\
    &\quad\left. +(s+l-2\xi)\Msq\StrCons{\beta}-(2s-2l-3)\S^2\StrCons{\gamma}\right)\Q{l-2j}{s-1-2j}{\xi-2j}{m+n}\\
    &\quad +\frac{l-\xi}{4}\S\left(2(4s(s-2)+4l^2-2\xi^2-l(10s-2\xi-11)+2s\xi-3\xi+5)\StrCons{\alpha}-4(2l-2\xi+1)\StrCons{\beta}\right.\\
    &\quad\left. -(l+1-\xi)(2s-2-l-\xi)^{\underline{2}}\StrCons{\gamma}\right)\Q{l-1-2j}{s-1-2j}{\xi-2j}{m+n}+\frac{(l-\xi)^{\underline{2}}(2s+1-l-\xi)^{\underline{2}}}{4}\StrCons{\beta}\Q{l-2-2j}{s-1-2j}{\xi-2j}{m+n}\\
    &\quad +(s-2-l)^{\underline{2}}\mathcal{M}^4\StrCons{\alpha}\Q{l+1-2j}{s-2-2j}{\xi-1-2j}{m+n}-(s-2-l)\Msq\S\left(2(s+\xi-2l-3)\StrCons{\alpha}\right.\\
    &\quad\left. -(l+1-\xi)(2s-l-\xi-3)\StrCons{\gamma}\right)\Q{l-2j}{s-2-2j}{\xi-1-2j}{m+n}+\frac{1}{4}\left(\left((2s-\xi-2)^{\underline{2}}\xi^{\underline{2}}\Msq-2l^3(2s-3)\Msq+l^4\Msq\right.\right.\\
    &\quad\left. +4(2s-3)(2\xi-1)\S^2+16l^2\S^2+l^2(4s^2-2\xi(\xi+2)+2s(2\xi-7)+11)\Msq-2l(8(s+\xi-2)\S^2\right.\\
    &\quad\left. +(2s-3)(s(2\xi-1)-\xi(\xi+2)+1)\Msq)\right)\StrCons{\alpha}+2\Msq\left(2s^2-l(2l^2+1)+l^2(4s+2\xi-3)\right.\\
    &\quad\left. -2l(4s-\xi-3)\xi+s(4\xi^{\underline{2}}-2)-\xi(2\xi^2+\xi-3)\right)\StrCons{\beta}+4\S^2(2s-2l-3)\left(s(2\xi-1)-\xi(\xi+2)-l(2s-3)\right.\\
    &\quad\left.\left. +l^2+1\right)\StrCons{\gamma}\right)\Q{l-1-2j}{s-2-2j}{\xi-1-2j}{m+n}-\frac{l-\xi}{2}\S\left((l-\xi-1)(2s-1-l-\xi)^{\underline{2}}\StrCons{\alpha}-2(2s-1-2\xi\right.\\
    &\quad\left.\left. -2(l-\xi)(2s-l-\xi))\StrCons{\beta}\right)\Q{l-2-2j}{s-2-2j}{\xi-1-2j}{m+n}\left. +\frac{(l-\xi)^{\underline{2}}(2s+1-l-\xi)^{\underline{2}}}{4}\StrCons{\beta}\Q{l-3-2j}{s-2-2j}{\xi-1-2j}{m+n}\right]\right\}\,.
\end{split}
\end{align}
\end{subequations}
\endgroup
}%
We note two special cases of the above rules:
\begin{subequations}\label{mult_Q_spin2_spec}
\begin{align}
\begin{split}
	P_n\star\Q{l}{s}{0}{m}&=\Q{l}{s+1}{0}{m+n}- l N^s_1(m,n)\left(\Q{l}{s+1}{1}{m+n}+\frac{2s+1-l}{2}\Q{l-1}{s}{0}{m+n}\right)\\
	&\quad +N_2^s(m,n)\left( l^{\underline{2}}\Q{l}{s+1}{2}{m+n}+\frac{l^{\underline{2}}(4s-2l-1)}{2}\Q{l-1}{s}{1}{m+n}\right.\\
	&\quad -(s-1-l)(s-1+l)\Msq \Q{l}{s-1}{0}{m+n}-l(2l-1)\S \Q{l-1}{s-1}{0}{m+n}\\
	&\quad \left. +\frac{l^{\underline{2}}(2s+1-l)^{\underline{2}}}{4}\Q{l-2}{s-1}{0}{m+n}\right)\,,
\end{split}\\
\begin{split}
	P_n\star\Q{l}{s}{1}{m}&=\Q{l}{s+1}{1}{m+n}+\Q{l-1}{s}{0}{m+n}-N_1^{s-1}(m,n)\left((l-1)\Q{l}{s+1}{2}{m+n} \vphantom{\frac{(l-1)(2s-l)}{2}}\right.\\
	&\quad +\frac{(l-1)(2s-l)}{2}\Q{l-1}{s}{1}{m+n}+l\Msq \Q{l}{s-1}{0}{m+n}-(2l-1)\S \Q{l-1}{s-1}{0}{m+n}\\
	&\quad \left.+\frac{(l-1)(2s-l)}{2}\Q{l-2}{s-1}{0}{m+n}\right)\\
	&\quad +N_2^{s-1}(m,n)\left((l-1)^{\underline{2}}\Q{l}{s+1}{3}{m+n}+\frac{(l-1)^{\underline{2}}(4s-2l-1)}{2}\Q{l-1}{s}{2}{m+n}\right.\\
	&\quad -(s-2-l)(s-2+l)\Msq \Q{l}{s-1}{1}{m+n}-(l-1)(2l-1)\S\Q{l-1}{s-1}{1}{m+n}\\
	&\quad +\frac{(l-1)^{\underline{2}}(2s-l)^{\underline{2}}}{4}\Q{l-2}{s-1}{1}{m+n}\\
	&\quad +\frac{2s^2+4ls(l-2)-2s-l(2l^2+l-7)}{2}\Msq\Q{l-1}{s-2}{0}{m+n}\\
	&\quad \left. +(l-1)(6s-2l(2s-l)-5)\S\Q{l-2}{s-2}{0}{m+n}+\frac{(l-1)^{\underline{2}}(2s-l)^{\underline{2}}}{4}\Q{l-3}{s-2}{0}{m+n}\right)\,.
\end{split}
\end{align}
\end{subequations}
Note that, in principle these relations contain all information about the complete set of product rules.
\section{Product Rules of the Right Slice \texorpdfstring{$\ihs^{\text{\tiny (R)}}(\Msq,\S)$}{ihs(R)(M\texttwosuperior,S)}}\label{app_strCons_right}
We will give a short outline on how to gain expressions for the multiplication (commutation) rules of the set of $(l=0)$- and $(l=1)$-generators. Let us start with the first case.

To get an expression for the product $\Q{0}{s}{0}{m}\star\Q{0}{t}{0}{n}$ one may begin with the easy relation
\begin{align}
    \Q{0}{s}{0}{m}\star P_n=\Q{0}{s+1}{0}{m+n}-\frac{\mathpzc{n}_2^s(m,n)}{4(s-\sfrac{1}{2})^{\underline{2}}}\Msq \Q{0}{s-1}{0}{m+n}\,,
\end{align}
where the short-hand (\ref{small_n}) is used. This product rule can be iterated to gain an expression for a product involving arbitrary powers of $P_{-1}$, yielding
\begin{align}
    \Q{0}{s}{0}{m}\star(P_{-1})^\sigma=\sum_{u=0}^{\sigma}\frac{(-1)^u\mathcal{M}^{2u}}{4^u}\frac{(s+m-1)^{\underline{2u}}}{(s-\sfrac{3}{2})^{\underline{u}}(s+\sigma-u-\sfrac{1}{2})^{\underline{u}}}\binom{\sigma}{u}\Q{0}{s+\sigma-2u}{0}{m-\sigma}\,.
\end{align}
Thus, we found an expression for the product of an arbitrary $(l=0)$-generator with a lowest-weight $(l=0)$-generator. One may convince oneself that such an expression can be generalized to all modes by replacing the $m$-dependent factorial by the function $\mathcal{N}^{st}_u(m,n)$ as defined in (\ref{lone-star_N}), using
\begin{align}
    (s+m-1)^{\underline{2u}}=\frac{\mathcal{N}^{st}_{2u}(m,-(t-1))}{4^u(t-1)^{\underline{u}}(t-\sfrac{3}{2})^{\underline{u}}}\,.
\end{align}
This leads us to the product rules
\begin{align}
    \Q{0}{s}{0}{m}\star\Q{0}{t}{0}{n}&=\sum_{u=0}^{\left\lfloor\!\frac{s+t-2}{2}\!\right\rfloor}\frac{(-1)^u\mathcal{M}^{2u}}{4^{2u}u!}\frac{\mathcal{N}^{st}_{2u}(m,n)}{(s-\sfrac{3}{2})^{\underline{u}}(t-\sfrac{3}{2})^{\underline{u}}(s+t-u-\sfrac{3}{2})^{\underline{u}}}\,\Q{0}{s+t-1-2u}{0}{m+n}\,.
\end{align}
The case $l=1$ appears to be more involved. However, we may employ a trick: The product rules (\ref{mult_Q_spin2}) enable us to re-write the $(l=1)$-generators in terms of $(l=0)$-generators as
\begin{subequations}\label{rewrite_generators}
\begin{align}
\begin{split}
    \Q{1}{s}{0}{m}&=\frac{(s+m-1)^{\underline{2}}}{4(s-1)(s-\sfrac{3}{2})}J_1\star\Q{0}{s-1}{0}{m-1}+\frac{(s+m-1)(s-m-1)}{2(s-1)(s-\sfrac{3}{2})}J_0\star\Q{0}{s-1}{0}{m}\\
    &\quad +\frac{(s-m-1)^{\underline{2}}}{4(s-1)(s-\sfrac{3}{2})}J_{-1}\star\Q{0}{s-1}{0}{m+1}\,,
\end{split}\\
\begin{split}
    \Q{1}{s}{1}{m}&=-\frac{m+s-2}{2(s-2)}J_1\star\Q{0}{s-1}{0}{m-1}+\frac{m}{s-2}J_0\star\Q{0}{s-1}{0}{m}-\frac{m-s+2}{2(s-2)}J_{-1}\star\Q{0}{s-1}{0}{m+1}\,.
\end{split}
\end{align}
\end{subequations}
This allows us to trace everything back to the structure constants of the $(l=0)$-slice. In a first step one may derive the respective lowest-weight products and commutators:
\begingroup
\allowdisplaybreaks
{\scriptsize
\begin{subequations}\label{spins_comm_acomm}
\begin{align}
\begin{split}
    \Q{1}{s}{0}{m}\star(P_{-1})^{\sigma}&=\frac{1}{s-1}\sum_{u=0}^{\sigma}\frac{(-1)^u\mathcal{M}^{2u}}{4^u (s-\sfrac{3}{2})^{\underline{u}}}\binom{\sigma}{u}\left(\frac{(s+m-1)^{\underline{2u}}}{(s+\sigma-u-\sfrac{1}{2})^{\underline{u}}}\frac{s(s+\sigma)-(2u+1)(s+\sigma-u)}{s+\sigma-2u}\Q{1}{s+\sigma-2u}{0}{m-\sigma}\right.\\
    &\left.\quad -\frac{(s+m-1)^{\underline{2u+1}}}{(s+\sigma-u-\sfrac{3}{2})^{\underline{u}}}\frac{\sigma-u}{s+\sigma-2u-1}\Q{1}{s+\sigma-2u}{1}{m-\sigma}+\frac{2(s+m-1)^{\underline{2u}}}{(s+\sigma-u-\sfrac{1}{2})^{\underline{u}}}\frac{u(\sigma-u+\sfrac{1}{2})}{s+\sigma-2u}\frac{\S}{\Msq}\Q{0}{s+\sigma-2u}{0}{m-\sigma}\right)\,,
\end{split}\\
\begin{split}
    (P_{-1})^{\sigma}\star\Q{1}{s}{0}{m}&=\Q{1}{s}{0}{m}\star(P_{-1})^{\sigma}-\sum_{u=0}^{\sigma}\frac{(-1)^u\mathcal{M}^{2u}}{4^u(s-1)}\frac{(s+m-1)^{\underline{2u+1}}(\sigma-u)}{(s-\sfrac{3}{2})^{\underline{u}}(s+\sigma-u-\sfrac{3}{2})^{\underline{u}}}\binom{\sigma}{u}\Q{0}{s+\sigma-1-2u}{0}{m-\sigma}\,,
\end{split}\\
\begin{split}
    \Q{1}{s}{1}{m}\star(P_{-1})^{\sigma}&=\frac{1}{s-2}\sum_{u=0}^{\sigma}\frac{(-1)^u\mathcal{M}^{2u}}{4^u}\binom{\sigma}{u}\left(\frac{(s+m-2)^{\underline{2u-1}}}{(s-\sfrac{5}{2})^{\underline{u-1}}(s+\sigma-u-\sfrac{3}{2})^{\underline{u-1}}}\frac{4u}{s+\sigma-2u}\Q{1}{s+\sigma-2u}{0}{m-\sigma}\right.\\
    &\quad +\frac{(s+m-2)^{\underline{2u}}}{(s-\sfrac{5}{2})^{\underline{u}}(s+\sigma-u-\sfrac{3}{2})^{\underline{u}}}\frac{s^2+2u^2+s(\sigma-2u-3)-2\sigma(u+1)+3u+2}{s+\sigma-2u-1}\Q{1}{s+\sigma-2u}{1}{m-\sigma}\\
    &\quad \left.+\frac{(s+m-2)^{\underline{2u+1}}}{(s-\sfrac{5}{2})^{\underline{u}}(s+\sigma-u-\sfrac{5}{2})^{\underline{u}}}\frac{\sigma-u}{s+\sigma-2u-2}\S\Q{0}{s+\sigma-2-2u}{0}{m-\sigma}\right)\,,
\end{split}\\
\begin{split}
    (P_{-1})^{\sigma}\star \Q{1}{s}{1}{m}&=\Q{1}{s}{1}{m}\star(P_{-1})^{\sigma}-\sum_{u=0}^{\sigma}\frac{(-1)^u\mathcal{M}^{2u}}{4^u(s-2)}\frac{(s+m-2)^{\underline{2u}}\left(2u(s-u-\sfrac{3}{2})-(s-2u-2)\sigma\right)}{(s-\sfrac{5}{2})^{\underline{u}}(s+\sigma-u-\sfrac{3}{2})^{\underline{u}}}\binom{\sigma}{u}\Q{0}{s+\sigma-1-2u}{0}{m-\sigma}\,,
\end{split}\\
\begin{split}
    \left[\Q{0}{s}{0}{m}\,,J_{-1}(P_{-1})^{\sigma-1}\right]&=\sum_{u=0}^{\sigma-1}\frac{(-1)^{u}\mathcal{M}^{2u}}{4^u}\frac{(s+m-1)^{\underline{2u+1}}}{(s-\sfrac{3}{2})^{\underline{u}}(s+\sigma-u-\sfrac{3}{2})^{\underline{u}}}\binom{\sigma-1}{u}\Q{0}{s+\sigma-1-2u}{0}{m-\sigma}\,,
\end{split}\\
\begin{split}
    \left[\Q{1}{s}{0}{m}\,,J_{-1}(P_{-1})^{\sigma-1}\right]&=\sum_{u=0}^{\sigma-1}\frac{(-1)^u\mathcal{M}^{2u}}{4^u(s-1)}\binom{\sigma-1}{u}\left(\frac{(s+m-1)^{\underline{2u+1}}(s+\sigma-2-2u)}{(s-\sfrac{3}{2})^{\underline{u}}(s+\sigma-u-\sfrac{3}{2})^{\underline{u}}}\Q{1}{s+\sigma-1-2u}{0}{m-\sigma}\right.\\
    &\quad \left.-\frac{\S(s+m-1)^{\underline{2u+3}}(\sigma-1-u)}{2(s-\sfrac{3}{2})^{\underline{u+1}}(s+\sigma-u-\sfrac{5}{2})^{\underline{u+1}}}\Q{0}{s+\sigma-3-2u}{0}{m-\sigma}\right)\,,
\end{split}\\
\begin{split}
    \left[\Q{1}{s}{1}{m}\,,J_{-1}(P_{-1})^{\sigma-1}\right]&=\sum_{u=0}^{\sigma-1}\frac{(-1)^u\mathcal{M}^{2u}}{4^u(s-2)}\binom{\sigma-1}{u}\left[\frac{(s+m-2)^{\underline{2u}}}{(s-\sfrac{5}{2})^{\underline{u}}(s+\sigma-u-\sfrac{5}{2})^{\underline{u}}}\Bigg(2u(s-u-\sfrac{3}{2})-(\sigma-1)(s-2u-2)\vphantom{\frac{4s(s-\sfrac{3}{2})}{2}}\right.\\
    &\quad\quad\quad\left. +\frac{4u(s-u-\sfrac{3}{2})(s+\sigma-2u-\sfrac{3}{2})}{s+\sigma-1-2u}\right)\Q{1}{s+\sigma-1-2u}{0}{m-\sigma}\\
    &\quad +\frac{(s+m-2)^{\underline{2u+1}}}{(s-\sfrac{5}{2})^{\underline{u}}(s+\sigma-u-\sfrac{5}{2})^{\underline{u}}}\frac{(s-1)^{\underline{2}}}{s+\sigma-2-2u}\Q{1}{s+\sigma-1-2u}{1}{m-\sigma}\\
    &\quad +\frac{\S(s+m-2)^{\underline{2u+2}}}{(s-\sfrac{5}{2})^{\underline{u}}(s+\sigma-u-\sfrac{5}{2})^{\underline{u+1}}}\frac{2\sigma^2+s(3\sigma-4u-3)-3\sigma(2u+3)+4u(u+3)+7}{2(s+\sigma-3-2u)}\Q{0}{s+\sigma-3-2u}{0}{m-\sigma}\\
    &\quad +\frac{\Msq(s+m-2)^{\underline{2u+2}}}{(s-\sfrac{5}{2})^{\underline{u}}(s+\sigma-u-\sfrac{5}{2})^{\underline{u+2}}}\frac{(s+\sigma-4-2u)\left(2u(s-u-\sfrac{3}{2})-(\sigma-1)(s-2-2u)\right)}{4(s+\sigma-3-2u)}\Q{1}{s+\sigma-3-2u}{0}{m-\sigma}\Bigg]\,.
\end{split}
\end{align}
\end{subequations}
}%
\endgroup
At this point we are in possession of all identities we need to evaluate the equations of motion in the main part of this paper. However, let us also note the generalized expressions for (some of) the above product rules and commutators; these are
\begingroup
\allowdisplaybreaks
{\scriptsize
\begin{subequations}
\begin{align}
\begin{split}
    \Q{1}{s}{0}{m}\star\Q{0}{t}{0}{n}&=\frac{1}{s-1}\sum_{u=0}^{\left\lfloor\!\frac{s+t-3}{2}\!\right\rfloor}\frac{(-1)^u\mathcal{M}^{2u}}{4^{2u}u!}\frac{\left((s+t-1)(s-1-2u)+u(2u+1)\right)\mathcal{N}^{st}_{2u}(m,n)}{(s-\sfrac{3}{2})^{\underline{u}}(t-\sfrac{3}{2})^{\underline{u}}(s+t-u-\sfrac{3}{2})^{\underline{u}}(s+t-2u-1)}\,\Q{1}{s+t-1-2u}{0}{m+n}\\
    &\quad -\frac{1}{2(s-1)}\sum_{u=0}^{\left\lfloor\!\frac{s+t-4}{2}\!\right\rfloor}\frac{(-1)^u\mathcal{M}^{2u}}{4^{2u}u!}\frac{\mathcal{N}^{st}_{2u+1}(m,n)}{(s-\sfrac{3}{2})^{\underline{u}}(t-\sfrac{3}{2})^{\underline{u}}(s+t-u-\sfrac{5}{2})^{\underline{u}}(s+t-2u-2)}\,\Q{1}{s+t-1-2u}{1}{m+n}\\
    &\quad -\frac{2\S}{s-1}\sum_{u=0}^{\left\lfloor\!\frac{s+t-4}{2}\!\right\rfloor}\frac{(-1)^u\mathcal{M}^{2u}}{4^{2u+2}u!}\frac{\mathcal{N}^{st}_{2u+2}(m,n)}{(s-\sfrac{3}{2})^{\underline{u+1}}(t-\sfrac{3}{2})^{\underline{u}}(s+t-u-\sfrac{5}{2})^{\underline{u+1}}(s+t-2u-3)}\,\Q{0}{s+t-3-2u}{0}{m+n}\,,
\end{split}\\
\begin{split}
    \Q{1}{s}{1}{m}\star\Q{0}{t}{0}{n}&=\frac{2}{s-2}\sum_{u=0}^{\left\lfloor\!\frac{s+t-3}{2}\!\right\rfloor}\frac{(-1)^u\mathcal{M}^{2u}}{4^{2u-1}u!}\frac{u\mathcal{N}^{s-1,t}_{2u-1}(m,n)}{(s-\sfrac{5}{2})^{\underline{u-1}}(t-\sfrac{3}{2})^{\underline{u-1}}(s+t-u-\sfrac{5}{2})^{\underline{u-1}}(s+t-2u-1)}\,\Q{1}{s+t-1-2u}{0}{m+n}\\
    &\quad +\frac{1}{s-2}\sum_{u=0}^{\left\lfloor\!\frac{s+t-4}{2}\!\right\rfloor}\frac{(-1)^u\mathcal{M}^{2u}}{4^{2u}u!}\frac{\left(s^2+2u^2+s(t-2u-4)-2(t-1)(u+1)+3u+2\right)\mathcal{N}^{s-1,t}_{2u}(m,n)}{(s-\sfrac{5}{2})^{\underline{u}}(t-\sfrac{3}{2})^{\underline{u}}(s+t-u-\sfrac{5}{2})^{\underline{u}}(s+t-2u-2)}\,\Q{1}{s+t-1-2u}{1}{m+n}\\
    &\quad +\frac{2\S}{s-2}\sum_{u=0}^{\left\lfloor\!\frac{s+t-4}{2}\!\right\rfloor}\frac{(-1)^u\mathcal{M}^{2u}}{4^{2u+1}u!}\frac{\mathcal{N}^{s-1,t}_{2u+1}(m,n)}{(s-\sfrac{5}{2})^{\underline{u}}(t-\sfrac{3}{2})^{\underline{u}}(s+t-u-\sfrac{7}{2})^{\underline{u}}(s+t-2u-3)}\,\Q{0}{s+t-3-2u}{0}{m+n}\,,
\end{split}\\
    \left[\Q{1}{s}{0}{m},\Q{0}{t}{0}{n}\right]&=\frac{1}{2(s-1)}\sum_{u=0}^{\left\lfloor\!\frac{s+t-3}{2}\!\right\rfloor}\frac{(-1)^u\mathcal{M}^{2u}}{4^{2u}u!}\frac{\mathcal{N}^{st}_{2u+1}(m,n)}{(s-\sfrac{3}{2})^{\underline{u}}(t-\sfrac{3}{2})^{\underline{u}}(s+t-u-\sfrac{5}{2})^{\underline{u}}}\Q{0}{s+t-2-2u}{0}{m+n}\,,\\
    \left[\Q{1}{s}{1}{m},\Q{0}{t}{0}{n}\right]&=\frac{1}{s-2}\sum_{u=0}^{\left\lfloor\!\frac{s+t-3}{2}\!\right\rfloor}\frac{(-1)^u\mathcal{M}^{2u}}{4^{2u}u!}\frac{\left(2u(s-u-\sfrac{3}{2})-(s-2u-2)(t-1)\right)\mathcal{N}^{s-1,t}_{2u}(m,n)}{(s-\sfrac{5}{2})^{\underline{u}}(t-\sfrac{3}{2})^{\underline{u}}(s+t-u-\sfrac{5}{2})^{\underline{u}}}\Q{0}{s+t-2-2u}{0}{m+n}\,,\\
\begin{split}    
    \left[\Q{1}{s}{0}{m},\Q{1}{t}{0}{n}\right]&=\frac{1}{2(s-1)(t-1)}\sum_{u=0}^{\left\lfloor\!\frac{s+t-4}{2}\!\right\rfloor}\frac{(-1)^u\mathcal{M}^{2u}}{4^{2u}u!}\frac{(s+t-3-2u)\mathcal{N}^{st}_{2u+1}(m,n)}{(s-\sfrac{3}{2})^{\underline{u}}(t-\sfrac{3}{2})^{\underline{u}}(s+t-u-\sfrac{5}{2})^{\underline{u}}}\Q{1}{s+t-2-2u}{0}{m+n}\\
    &\quad -\frac{\S}{(s-1)(t-1)}\sum_{u=0}^{\left\lfloor\!\frac{s+t-5}{2}\!\right\rfloor}\frac{(-1)^u\mathcal{M}^{2u}}{4^{2u+2}u!}\frac{\mathcal{N}^{st}_{2u+3}(m,n)}{(s-\sfrac{3}{2})^{\underline{u+1}}(t-\sfrac{3}{2})^{\underline{u+1}}(s+t-u-\sfrac{7}{2})^{\underline{u+1}}}\Q{0}{s+t-4-2u}{0}{m+n}\,.
\end{split}
\end{align}
\end{subequations}
}%
\endgroup

We would like to give a final remark on the derivation of product rules: The expansion (\ref{rewrite_generators}) actually turns out to be a special case of the fact that any $\ihs$-generator can be written as a sum of products of highest-$l$ and $(l=0)$-generators, which could be helpful for the construction of additional product rules. It simply follows from the expansion (\ref{descendant_expansion}) together with a splitting of highest-weight generators,
\begin{align}
    \Q{l}{s}{\xi}{s-1-\xi}=\begin{cases}
        \Q{l}{l+1}{\xi}{l-\xi}\star\Q{0}{s-l}{0}{s-1-l}\,, & \xi\text{ even}\\[0.2cm]
        \Q{l}{l+2}{\xi}{l-\xi+1}\star\Q{0}{s-1-l}{0}{s-2-l}\,, & \xi\text{ odd}
    \end{cases}\ \ ,
\end{align}
and the expressions
\begin{subequations}
\begin{align}
    \Q{l}{s}{\xi}{m}\star (J_n)^k&=\sum_{j=0}^{k}\binom{k}{j}\left(m-(s-1-\xi)n\right)^{\overline{k-j,n}} (J_n)^j\star\Q{l}{s}{\xi}{m+(k-j)n}\,,\\
    (J_n)^k\star\Q{l}{s}{\xi}{m}&=\sum_{j=0}^{k}(-1)^{k-j}\binom{k}{j}\left(m-(s-1-\xi)n\right)^{\overline{k-j,n}} \Q{l}{s}{\xi}{m+(k-j)n}\star (J_n)^j\,,
\end{align}
\end{subequations}
which can easily be shown by induction from (\ref{adjointJ}). Here, $(a)^{\overline{k,n}}$ denotes the $n$-step rising factorial. The decomposition of generators takes the form
\begin{subequations}
\begin{align}
\begin{split}
    \Q{l}{s}{\xi}{m}\at{4}{1.5}{\xi\text{ even}}&=\frac{(s+m-\xi-1)!(s-m-\xi-1)!}{(2s-2\xi-2)!}\times\\
    &\quad \times\sum_{k=0}^{s-1-\xi-m}\binom{2l-2\xi}{k}\binom{2s-2l-2}{s-1-\xi-m-k}\Q{l}{l+1}{\xi}{l-\xi-k}\star\Q{0}{s-l}{0}{m-l+\xi+k}\,,
\end{split}\\
\begin{split}
    \Q{l}{s}{\xi}{m}\at{4}{1.5}{\xi\text{ odd}}&=\frac{(s+m-\xi-1)!(s-m-\xi-1)!}{(2s-2\xi-2)!}\times\\
    &\quad \times\sum_{k=0}^{s-1-\xi-m}\binom{2l-2\xi+2}{k}\binom{2s-2l-4}{s-1-\xi-m-k}\Q{l}{l+2}{\xi}{l+1-\xi-k}\star\Q{0}{s-1-l}{0}{m-1-l+\xi+k}\,.
\end{split}
\end{align}
\end{subequations}
\section{Equations of Motion}\label{app_eom}
In this section we would like to display sets of equations of motion we found for the coefficients $\c{l}{s}{\xi}{m}(u,r,\phi)$ in the master field $C$. We use the convention that coefficients with index combinations outside the allowed ranges are to be identified with zero, e.g. $\c{1}{2}{1}{m}\equiv 0$.

\subsection{Spin 2}\label{appsub_spin2}
Coordinate dependencies of the fields are suppressed for better readability, $\c{l}{s}{\xi}{m}\equiv\c{l}{s}{\xi}{m}(u,r,\phi)$.
\begingroup
\allowdisplaybreaks
\begin{subequations}
\begin{align}
\begin{split}
	0&=\partial_{\!u} \c{0}{s}{0}{m}+2\c{0}{s-1}{0}{m-1}-\frac{M(\phi)}{2}\c{0}{s-1}{0}{m+1}+\c{1}{s}{1}{m-1}-\frac{M(\phi)}{4}\c{1}{s}{1}{m+1}+\frac{s-m}{s^{\underline{2}}}\left(s \c{1}{s}{0}{m-1}-2\S\c{1}{s+1}{1}{m-1}\right)\\
	&\quad +\frac{(s+m)M(\phi)}{4s^{\underline{2}}}\left(s \c{1}{s}{0}{m+1}-2\S\c{1}{s+1}{1}{m+1}\right)+\frac{(s-m+1)^{\underline{2}}}{4s^2(s+\sfrac{1}{2})^{\underline{2}}}\left((s+1)^{\underline{2}}\Msq \c{1}{s+2}{1}{m-1}-2\S\c{1}{s+1}{0}{m-1}\right.\\
	&\quad \left.-2s^2\Msq\c{0}{s+1}{0}{m-1}\right)-\frac{(s+m+1)^{\underline{2}}M(\phi)}{16s^{\underline{2}}(s+\sfrac{1}{2})^{\underline{2}}}\left((s+1)^{\underline{2}}\Msq\c{1}{s+2}{1}{m+1}-2\S\c{1}{s+1}{0}{m+1}-2s^2\Msq\c{0}{s+1}{0}{m+1}\right)\,,
\end{split}\\
\begin{split}
	0&=\partial_{\!u} \c{1}{s}{0}{m}+2\c{1}{s-1}{0}{m-1}-\frac{M(\phi)}{2}\c{1}{s-1}{0}{m+1}+\frac{2(s-m)}{s^{\underline{2}}}\Msq \c{1}{s+1}{1}{m-1}+\frac{(s+m)M(\phi)}{2s^{\underline{2}}}\Msq\c{1}{s+1}{1}{m+1}\\
	&\quad -\frac{(s-m+1)^{\underline{2}}(s^2-1)}{2s^2(s+\sfrac{1}{2})^{\underline{2}}}\Msq\c{1}{s+1}{0}{m-1}+\frac{(s+m+1)^{\underline{2}}(s^2-1)M(\phi)}{8s^2(s+\sfrac{1}{2})^{\underline{2}}}\Msq\c{1}{s+1}{0}{m+1}\,,
\end{split}\\
\begin{split}
	0&=\partial_{\!u} \c{1}{s}{1}{m}+2\c{1}{s-1}{1}{m-1}-\frac{M(\phi)}{2}\c{1}{s-1}{1}{m+1}+\frac{2(s-m-1)}{(s-1)^{\underline{2}}} \c{1}{s-1}{0}{m-1}+\frac{(s+m-1)M(\phi)}{2(s-1)^{\underline{2}}}\c{1}{s-1}{0}{m+1}\\
	&\quad -\frac{(s-m)^{\underline{2}}s(s-2)}{2(s-1)^2(s-\sfrac{1}{2})^{\underline{2}}}\Msq\c{1}{s+1}{1}{m-1}+\frac{(s+m)^{\underline{2}}s(s-2)M(\phi)}{8(s-1)^2(s-\sfrac{1}{2})^{\underline{2}}}\Msq\c{1}{s+1}{1}{m+1}\,;
\end{split}\\
\begin{split}
	0&=\partial_{\!r} \c{0}{s}{0}{m}+\c{0}{s-1}{0}{m+1}+\frac{1}{2}\c{1}{s}{1}{m+1}-\frac{s+m}{s^{\underline{2}}}\left(\frac{s}{2}\c{1}{s}{0}{m+1}-\S\c{1}{s+1}{1}{m+1}\right)\\
	&\quad +\frac{(s+m+1)^{\underline{2}}}{4s^2(s+\sfrac{1}{2})^{\underline{2}}}\left(\frac{(s+1)^{\underline{2}}}{2}\Msq\c{1}{s+2}{1}{m+1}-\S\c{1}{s+1}{0}{m+1}-s^2\Msq\c{0}{s+1}{0}{m+1}\right)\,,
\end{split}\\
\begin{split}
	0&=\partial_{\!r} \c{1}{s}{0}{m}+\c{1}{s-1}{0}{m+1}-\frac{s+m}{s^{\underline{2}}}\Msq\c{1}{s+1}{1}{m+1}-\frac{(s+m+1)^{\underline{2}}(s^2-1)}{4s^2(s+\sfrac{1}{2})^{\underline{2}}}\Msq\c{1}{s+1}{0}{m+1}\,,
\end{split}\\
\begin{split}
	0&=\partial_{\!r} \c{1}{s}{1}{m}+\c{1}{s-1}{1}{m+1}-\frac{s+m-1}{(s-1)^{\underline{2}}}\c{1}{s-1}{0}{m+1}-\frac{(s+m)^{\underline{2}}s(s-2)}{4(s-1)^2(s-\sfrac{1}{2})^{\underline{2}}}\Msq\c{1}{s+1}{1}{m+1}\,;
\end{split}\\
\begin{split}
	0&=\partial_{\!\phi} \c{0}{s}{0}{m}+2r\c{0}{s-1}{0}{m}+r\c{1}{s}{1}{m}-N(u,\phi)\c{0}{s-1}{0}{m+1}-\frac{N(u,\phi)}{2}\c{1}{s}{1}{m+1}-\frac{m r}{s^{\underline{2}}}\left(s \c{1}{s}{0}{m}-2\S\c{1}{s+1}{1}{m}\right)\\
	&\quad +(s-m)\c{0}{s}{0}{m-1}+\frac{(s+m)M(\phi)}{4}\c{0}{s}{0}{m+1}+\frac{(s+m)N(u,\phi)}{2s^{\underline{2}}}\left(s \c{1}{s}{0}{m+1}-2\S\c{1}{s+1}{1}{m+1}\right)\\
	&\quad +\frac{(s+m)(s-m)r}{4s^2(s+\sfrac{1}{2})^{\underline{2}}}\left(2s^2\Msq\c{0}{s+1}{0}{m}+2\S\c{1}{s+1}{0}{m}-(s+1)^{\underline{2}}\Msq\c{1}{s+2}{1}{m}\right)\\
	&\quad +\frac{(s+m+1)^{\underline{2}}N(u,\phi)}{8s^2(s+\sfrac{1}{2})^{\underline{2}}}\left(2s^2\Msq\c{0}{s+1}{0}{m+1}+2\S\c{1}{s+1}{0}{m+1}-(s+1)^{\underline{2}}\Msq\c{1}{s+2}{1}{m+1}\right)\,,
\end{split}\\
\begin{split}
	0&=\partial_{\!\phi} \c{1}{s}{0}{m}+2r\c{1}{s-1}{0}{m}-N(u,\phi)\c{1}{s-1}{0}{m+1}-\frac{2mr}{s^{\underline{2}}}\Msq\c{1}{s+1}{1}{m}+(s-m)\c{1}{s}{0}{m-1}\\
	&\quad +\frac{(s+m)M(\phi)}{4}\c{1}{s}{0}{m+1}+\frac{(s+m)N(u,\phi)}{s^{\underline{2}}}\Msq\c{1}{s+1}{1}{m+1}\\
	&\quad +\frac{(s+m)(s-m)(s^2-1)r}{2s^2(s+\sfrac{1}{2})^{\underline{2}}}\Msq\c{1}{s+1}{0}{m}+\frac{(s+m+1)^{\underline{2}}(s^2-1)N(u,\phi)}{4s^2(s+\sfrac{1}{2})^{\underline{2}}}\Msq\c{1}{s+1}{0}{m+1}\,,
\end{split}\\
\begin{split}
	0&=\partial_{\!\phi} \c{1}{s}{1}{m}+2r\c{1}{s-1}{1}{m}-N(u,\phi)\c{1}{s-1}{1}{m+1}-\frac{2mr}{(s-1)^{\underline{2}}}\c{1}{s-1}{0}{m}+(s-m-1)\c{1}{s}{1}{m-1}\\
	&\quad +\frac{(s+m-1)M(\phi)}{4}\c{1}{s}{1}{m+1}+\frac{(s+m-1)N(u,\phi)}{(s-1)^{\underline{2}}}\c{1}{s-1}{0}{m+1}\\
	&\quad +\frac{(s+m-1)(s-m-1)s(s-2)r}{2(s-1)^2(s-\sfrac{1}{2})^{\underline{2}}}\Msq\c{1}{s+1}{1}{m}+\frac{(s+m)^{\underline{2}}s(s-2)N(u,\phi)}{4(s-1)^2(s-\sfrac{1}{2})^{\underline{2}}}\Msq\c{1}{s+1}{1}{m+1}\,.
\end{split}
\end{align}
\end{subequations}
\endgroup
\subsection{Spin \texorpdfstring{$s$}{s}}\label{appsub_spins}
The abbreviations $N_1^s(m,n)$ and $N_2^s(m,n)$ are defined in equations (\ref{def_capitalN}). Coordinate dependencies are suppressed for better readability, $\c{l}{s}{\xi}{m}\equiv\c{l}{s}{\xi}{m}(u,r,\phi)$ and $N\equiv N(u,\phi)$.
\begingroup
\allowdisplaybreaks
\begin{subequations}
\begin{align}
    \begin{split}
    0&=\partial_{\!u}\c{0}{s}{0}{m}+2\c{0}{s-1}{0}{m-1}-2s^2 N_2^{s+1}(m-1,1)\Msq\c{0}{s+1}{0}{m-1}-s N_1^s(m-1,1)\c{1}{s}{0}{m-1}+\c{1}{s}{1}{m-1}\\
    &\quad+(s+1)^{\underline{2}}N_2^{s+1}(m-1,1)\Msq\c{1}{s+2}{1}{m-1}-\sum_{\sigma=2}^{\tilde{s}}\sum_{k=0}^{\sigma-1}\frac{(-1)^k\mathcal{M}^{2k}Z^{(\sigma)}}{4^{k+1}}\binom{\sigma-1}{k}\left[\vphantom{\frac{(a)^{\underline{k}}}{(b)^{\underline{k}}}}\right.\\
    &\quad\quad \frac{2(s+m+2k-1)^{\underline{2k}}}{(s+2k-\sigma-\sfrac{1}{2})^{\underline{k}}(s+k-\sfrac{1}{2})^{k}}\c{0}{s+2k+1-\sigma}{0}{m+\sigma-1}\\
    &\quad\quad -\frac{(s+m+2k)^{\underline{2k+1}}(\sigma-1-k)}{(s+2k+1-\sigma)(s+2k-\sigma+\sfrac{1}{2})^{\underline{k}}(s+k-\sfrac{1}{2})^{\underline{k}}}\c{1}{s+2k+2-\sigma}{0}{m+\sigma-1}\\
    &\quad\quad \left.-\frac{(s+m+2k-1)^{\underline{2k}}\left(2k(s+k-\sigma+\sfrac{1}{2})-(s-\sigma)(\sigma-1)\right)}{(s+2k-\sigma)(s+2k-\sigma-\sfrac{1}{2})^{\underline{k}}(s+k-\sfrac{1}{2})^{\underline{k}}}\c{1}{s+2k+2-\sigma}{1}{m+\sigma-1}\right]\,,
    \end{split}\\
    \begin{split}
    0&=\partial_{\!u}\c{1}{s}{0}{m}+2\c{1}{s-1}{0}{m-1}-2(s^2-1)N_2^{s+1}(m-1,1)\Msq\c{1}{s+1}{0}{m-1}-2N_1^s(m-1,1)\Msq\c{1}{s+1}{1}{m-1}\\
     &\quad -\sum_{\sigma=2}^{\tilde{s}}\sum_{k=0}^{\sigma-1}\frac{(-1)^k\mathcal{M}^{2k}Z^{(\sigma)}}{4^{k+1}}\binom{\sigma-1}{k}\left[\vphantom{\frac{(a)^{\underline{k}}}{(b)^{\underline{k}}}}\right.\\
     &\quad\quad \frac{2(s+m+2k-1)^{\underline{2k}}\left((s+2k+1-\sigma)(s+2k)-(2k+1)(s+k)\right)}{(s+2k-\sigma)(s+2k-\sigma-\sfrac{1}{2})^{\underline{k}}(s+k-\sfrac{1}{2})^{\underline{k}}s}\c{1}{s+2k+1-\sigma}{0}{m+\sigma-1}\\
     &\quad\quad\left. +\frac{8(s+m+2k-2)^{\underline{2k-1}}k}{(s+2k-1-\sigma)(s+2k-\sigma-\sfrac{3}{2})^{\underline{k-1}}(s+k-\sfrac{3}{2})^{\underline{k-1}}s}\c{1}{s+2k+1-\sigma}{1}{m+\sigma-1}\right]\,,
    \end{split}\\
    \begin{split}
    0&=\partial_{\!u}\c{1}{s}{1}{m}-2N_1^{s-1}(m-1,1)\c{1}{s-1}{0}{m-1}+2\c{1}{s-1}{1}{m-1}-2s(s-2)N_2^{s}(m-1,1)\Msq\c{1}{s+1}{1}{m-1}\\
    &\quad +\sum_{\sigma=2}^{\tilde{s}}\sum_{k=0}^{\sigma-1}\frac{(-1)^k\mathcal{M}^{2k}Z^{(\sigma)}}{4^{k+1}}\binom{\sigma-1}{k}\left[\vphantom{\frac{(a)^{\underline{k}}}{(b)^{\underline{k}}}}\right.\\
    &\quad\quad \frac{2(s+m+2k-1)^{\underline{2k+1}}(\sigma-1-k)}{(s+2k-\sigma)(s+2k-\sigma-\sfrac{1}{2})^{\underline{k}}(s+k-\sfrac{3}{2})^{\underline{k}}(s-1)}\c{1}{s+2k+1-\sigma}{0}{m+\sigma-1}\\
    &\quad\quad \left. -\frac{2(s+m+2k-2)^{\underline{2k}}\left(2k^2+2k(s-\sigma-\sfrac{1}{2})+(s-1)(s-1-\sigma)\right)}{(s+2k-\sigma-1)(s+2k-\sigma-\sfrac{3}{2})^{\underline{k}}(s+k-\sfrac{3}{2})^{\underline{k}}(s-1)}\c{1}{s+2k+1-\sigma}{1}{m+\sigma-1}\right]\,;
    \end{split}\\[0.4cm]
    \begin{split}
    0&=\partial_{\!r}\c{0}{s}{0}{m}+\c{0}{s-1}{0}{m+1}-s^2N_2^{s+1}(m+1,-1)\Msq\c{0}{s+1}{0}{m+1}-\frac{s}{2}N_1^s(m+1,-1)\c{1}{s}{0}{m+1}+\frac{1}{2}\c{1}{s}{1}{m+1}\\
    &\quad +\frac{(s+1)^{\underline{2}}}{2}N_2^{s+1}(m+1,-1)\Msq\c{1}{s+2}{1}{m+1}\,,
    \end{split}\\
    \begin{split}
    0&=\partial_{\!r}\c{1}{s}{0}{m}+\c{1}{s-1}{0}{m+1}-(s^2-1)N_2^{s+1}(m+1,-1)\Msq\c{1}{s+1}{0}{m+1}-N_1^s(m+1,-1)\Msq\c{1}{s+1}{1}{m+1}\,,
    \end{split}\\
    \begin{split}
    0&=\partial_{\!r}\c{1}{s}{1}{m}+\c{1}{s-1}{1}{m+1}-N_1^{s-1}(m+1,-1)\c{1}{s-1}{0}{m+1}-s(s-2)N_2^{s}(m+1,-1)\Msq\c{1}{s+1}{1}{m+1}\,;
    \end{split}\\[0.4cm]
    \begin{split}
    0&=\partial_{\!\phi}\c{0}{s}{0}{m}+(s-m)\c{0}{s}{0}{m-1}+2r\c{0}{s-1}{0}{m} -2r s^2 N_2^{s+1}(m,0)\Msq\c{0}{s+1}{0}{m}-rsN_1^s(m,0)\c{1}{s}{0}{m}\\
    &\quad +r\c{1}{s}{1}{m}+r(s+1)^{\underline{2}}N_2^{s+1}(m,0)\Msq \c{1}{s+2}{1}{m}-N\c{0}{s-1}{0}{m+1}+N s^2 N_2^{s+1}(m+1,-1)\Msq\c{0}{s+1}{0}{m+1}\\
    &\quad +\frac{Ns}{2}N_1^s(m+1,-1)\c{1}{s}{0}{m+1}-\frac{N}{2}\c{1}{s}{1}{m+1}-\frac{N(s+1)^{\underline{2}}}{2}N_2^{s+1}(m+1,-1)\Msq\c{1}{s+2}{1}{m+1}\\
    &\quad +\sum_{\sigma=2}^{\tilde{s}}\sum_{k=0}^{\sigma-2}\frac{(-1)^k(\sigma-1)\mathcal{M}^{2k}Z^{(\sigma)}}{4^{k+1}}\binom{\sigma-2}{k}\frac{(s+m+2k)^{\underline{2k+1}}}{(s+2k-\sigma+\sfrac{1}{2})^{\underline{k}}(s+k-\sfrac{1}{2})^{\underline{k}}}\c{0}{s+2k+2-\sigma}{0}{m+\sigma-1}\,,
    \end{split}\\
    \begin{split}
    0&=\partial_{\!\phi}\c{1}{s}{0}{m}+(s-m)\c{1}{s}{0}{m-1}+2r\c{1}{s-1}{0}{m}-2r(s^2-1)N_2^{s+1}(m,0)\Msq\c{1}{s+1}{0}{m}\\
    &\quad -2rN_1^s(m,0)\Msq\c{1}{s+1}{1}{m}-N\c{1}{s-1}{0}{m+1}+N(s^2-1)N_2^{s+1}(m+1,-1)\Msq\c{1}{s+1}{0}{m+1}\\
    &\quad +NN_1^s(m+1,-1)\Msq\c{1}{s+1}{1}{m+1}+\sum_{\sigma=2}^{\tilde{s}}\sum_{k=0}^{\sigma-2}\frac{(-1)^k(\sigma-1)\mathcal{M}^{2k}Z^{(\sigma)}}{4^{k+1}}\binom{\sigma-2}{k}\left[\vphantom{\frac{(a)^{\underline{k}}}{(b)^{\underline{k}}}}\right.\\
    &\quad\quad \frac{(s+m+2k)^{\underline{2k+1}}(s-1)}{(s+2k+1-\sigma)(s+2k-\sigma+\sfrac{1}{2})^{\underline{k}}(s+k-\sfrac{1}{2})^{\underline{k}}}\c{1}{s+2k+2-\sigma}{0}{m+\sigma-1}\\
    &\quad\quad +\frac{(s+m+2k-1)^{\underline{2k}}}{(s+2k-\sigma)(s+2k-\sigma-\sfrac{1}{2})^{\underline{k}}(s+k-\sfrac{3}{2})^{\underline{k}}}\left(2k(s+k-\sigma+\sfrac{1}{2}) \vphantom{\frac{(a)}{(b)}}\right.\\
    &\quad\quad\quad \left.-(\sigma-2)(s-\sigma)+\frac{4k(s+k-\sigma+\sfrac{1}{2})(s-\sfrac{1}{2})}{s}\right)\c{1}{s+2k+2-\sigma}{1}{m+\sigma-1}\\
    &\quad\quad +\frac{(s+m+2k+1)^{\underline{2k+2}}(s-1)}{4(s+2k+2-\sigma)(s+2k-\sigma+\sfrac{3}{2})^{\underline{k}}(s+k+\sfrac{1}{2})^{\underline{k+2}}s}\left(2k(s+k-\sigma+\sfrac{5}{2})\right.\\
    &\quad\quad\quad \left.\left.-(\sigma-2)(s+2-\sigma)\right)\Msq\c{1}{s+2k+4-\sigma}{1}{m+\sigma-1}\right]\,,
    \end{split}\\
    \begin{split}
    0&=\partial_{\!\phi}\c{1}{s}{1}{m}+(s-m-1)\c{1}{s}{1}{m-1}-2rN_1^{s-1}(m,0)\c{1}{s-1}{0}{m}+2r\c{1}{s-1}{1}{m}\\
    &\quad -2rs(s-2)N_2^s(m,0)\Msq\c{1}{s+1}{1}{m}+NN_1^{s-1}(m+1,-1)\c{1}{s-1}{0}{m+1}-N\c{1}{s-1}{1}{m+1}\\
    &\quad +Ns(s-2)N_2^s(m+1,-1)\Msq\c{1}{s+1}{1}{m+1}+\sum_{\sigma=2}^{\tilde{s}}\sum_{k=0}^{\sigma-2}\frac{(-1)^k(\sigma-1)\mathcal{M}^{2k}Z^{(\sigma)}}{4^{k+1}}\binom{\sigma-2}{k}\times\\
    &\quad\quad \times\frac{(s+m+2k-1)^{\underline{2k+1}}(s+2k+1-\sigma)^{\underline{2}}}{(s+2k-\sigma)(s+2k-\sigma-\sfrac{1}{2})^{\underline{k}}(s+k-\sfrac{3}{2})^{\underline{k}}(s-1)}\c{1}{s+2k+2-\sigma}{1}{m+\sigma-1}\,.
    \end{split}
\end{align}
\end{subequations}
\endgroup
\newpage
\printbibliography
%\bibliography{bibliography}
\end{document}